%% file: main.tex
\documentclass[a4paper,fleqn]{cas-dc}

\usepackage[T1]{fontenc}
\usepackage{amsmath}
\usepackage{amssymb}

\usepackage{graphicx,verbatim}

\usepackage{hyperref}
\hypersetup{
    colorlinks = true,
    allcolors  = blue
}
\usepackage{color}

\usepackage{bookmark}
\usepackage{booktabs}
\usepackage{csquotes}
\usepackage{cleveref}
\usepackage{subcaption}
\usepackage{siunitx}
\usepackage{threeparttable}
\usepackage[authoryear,longnamesfirst]{natbib}
\usepackage{algorithm}
\usepackage{algpseudocode}
\usepackage{rotating} 
\usepackage[toc,acronym,nomain,nopostdot,nonumberlist]{glossaries}
\setacronymstyle{long-short}
\loadglsentries{glossary.tex} 		
\glsdisablehyper

\newcommand{\CPI}{\textsc{cpi}} 
\newcommand{\CP}{\textsc{cp}} 
\newcommand{\CD}{\textsc{cd}} 
\newcommand{\CARD}[1]{\left\lvert #1 \right\rvert}
\newcommand{\ToneWoneMM}{\qtyproduct[product-units = power]{1x1x1}{\milli\meter} \gls{t1w}}
\newcommand{\FLAIRfiveMM}{\qtyproduct[product-units = power]{1x1x5}{\milli\meter} \gls{flair}}

\begin{document}

\let\WriteBookmarks\relax
\def\floatpagepagefraction{1}
\def\textpagefraction{.001}

\shorttitle{Uncertainty-Aware Cortical Surface Reconstruction}

\shortauthors{JD Nielsen et~al.}

\title [mode = title]{Uncertainty-Aware Cortical Surface Reconstruction and Spherical Registration for Heterogeneous Brain MRI}                      



\author[1,2]{Jesper Duemose Nielsen}[orcid=0000-0003-3697-1012]
\ead{jdue@dtu.dk}
\credit{Conceptualization, Methodology, Software, Visualization, Writing - Original Draft}
\cormark[1] 

\author[3]{Karthik Gopinath}
\credit{Methodology, Data Curation, Writing - Review and Editing}

\author[3,4]{Andrew Hoopes}
\credit{Methodology, Software, Writing - Review and Editing}

\author[3,4,5]{Adrian Dalca}
\credit{Methodology, Writing - Review and Editing}

\author[1,2]{Axel Thielscher}
\credit{Funding acquisition, Writing - Review and Editing}

\author[3,4,5,6]{Juan Eugenio Iglesias}
\credit{Methodology, Funding acquisition, Data Curation, Writing - Review and Editing}

\author[1,2,3]{Oula Puonti}
\credit{Methodology, Conceptualization, Data Curation, Funding acquisition, Writing - Original Draft}


\affiliation[1]{organization={Danish Research Centre for Magnetic Resonance, Department of Radiology and Nuclear Medicine, Copenhagen University Hospital -- Amager and Hvidovre},
    city={Copenhagen},
    country={Denmark}}
\affiliation[2]{organization={Department of Health Technology, Technical University of Denmark},
    city={Kgs. Lyngby},
    country={Denmark}}
\affiliation[3]{organization={Athinoula A. Martinos Center for Biomedical Imaging, Massachusetts General Hospital},
    city={Charlestown},
    state={MA},
    country={USA}}
\affiliation[4]{organization={Computer Science and Artificial Intelligence Lab, Massachusetts Institute of Technology},
    city={Cambridge},
    state={MA},
    country={USA}}
\affiliation[5]{organization={Department of Radiology, Harvard Medical School},
    city={Boston},
    state={MA},
    country={USA}}
\affiliation[6]{organization={Hawkes Institute, University College London},
    city={London},
    country={UK}}

\cortext[cor1]{Corresponding author}



\begin{abstract}
Surface-based cortical analysis is useful in many applications as it is a natural way to model the highly folded structure of the cerebral cortex, allowing quantities such as curvature and thickness to be easily calculated. Existing \gls{dl} methods have focused on \ToneWoneMM{} scans, thus replicating classical pipelines. Here we present a neural network that is trained exclusively on synthetic scans with random contrast and resolution. Consequently, it can be applied to almost any kind of \gls{mri} scan. In addition, our model provides vertex-wise uncertainty estimates of the predicted surface as well as a spherical registration to \emph{fsaverage} space. Our model performs on par with other state-of-the-art \gls{dl} methods on \ToneWoneMM{} scans and outperforms the only other method capable of estimating cortical surfaces from low resolution, non-\gls{t1w} scans.
The predicted uncertainty, moreover, correlates with overall scan quality on low resolution scans and highlights areas of poor surface reconstruction on \ToneWoneMM{} scans, suggesting that it can be used for quality control. Finally, the predicted spherical registrations prove comparable to other \gls{dl} and classical registration methods. These qualities, when combined with the fast inference time, will provide the ability to analyze large, existing databases of clinical \gls{mri} scans, with the potential for finding disease-specific patterns that would otherwise be infeasible in a research setting.
\end{abstract}


\begin{highlights}
\item A deep learning model reconstructs cortical surfaces from MRI of varying resolution and contrast, with estimated surface placement uncertainty.
\item Reconstruction accuracy matches other deep learning tools on high-quality data and exceeds the state-of-the-art on low-resolution clinical scans.
\item Uncertainty estimates track local prediction errors on high-quality data and overall image quality on clinical scans.
\item The model also yields a spherical registration to \emph{fsaverage}, matching state-of-the-art accuracy on both scan types.
\end{highlights}

\begin{keywords}
\sep Cortical surface reconstruction \sep Clinical data \sep Deep learning \sep Magnetic resonance imaging \sep Uncertainty
\end{keywords}

\maketitle

\glsresetall

\section{Introduction}

Surface-based analysis has become a central framework for studying the structure and function of the human cortex. Reconstruction of the \gls{wm} and \gls{gm} (pial) surfaces provides an anatomically informed coordinate system for representing and analyzing functional \gls{mri} data, and enables morphometric quantification of the cortical sheet. Seemingly simple morphometric quantities, such as the thickness, curvature, or surface area, have provided profound insights into cortical development in health and disease. For example, one representative result is the link between cortical thinning and dementia, which is strong enough that cortical thickness estimates can be used to predict conversion from mild cognitive impairment to Alzheimer's disease with a 76\% accuracy~\cite{querbes2009}. Localized changes in thickness, curvature and surface area have also been linked to autism spectrum disorder~\cite{mensen2017development, ecker2013brain}, schizophrenia~\cite{sun2023meta} and temporal lobe epilepsy~\cite{kemmotsu2011mri}. The cortical geometry is also linked to the underlying microstructural properties, including cytoarchitecture, myeloarchitecture and intracortical lamination~\cite{fischl2008cortical, paquola2026intracortical}.

Despite the usefulness of morphometric measures, current surface reconstruction tools are primarily tailored to high-quality, research-grade MRI scans with \gls{t1w} contrast. This limits their applicability to heterogeneous datasets, such as clinical MRI data, which typically have thick slices and do not include a high-resolution, isotropic \gls{t1w} scan, or \emph{ex vivo} acquisitions, which typically deviate significantly from standard \gls{t1w} and clinical scans in terms of resolution, contrast, and spatial coverage. Extending cortical analysis to these types of datasets would, on the one hand, enable large-scale studies of how cortical measures relate to disease symptoms using already existing clinical data, and, on the other hand, allow zooming into the connection between cortical geometry and microstructure using high-resolution \emph{ex vivo} data. In this paper, we aim to widen the scope of surface-based cortical analysis by introducing a model that can reconstruct surfaces from diverse MRI data and provides automated uncertainty estimates to highlight potential reconstruction errors.

\paragraph{Surface reconstruction of high-quality \gls{mri} scans.}
Classical surface reconstruction approaches such as FreeSurfer~\cite{fischl2012}, BrainSuite~\cite{shattuck2002}, Caret~\cite{vanessen2001}, BrainVoyager~\cite{goebel2012}, CAT~\cite{dahnke2013cortical}, and Civet~\cite{macdonald2000} rely on iterative optimization and geometry processing algorithms with carefully tuned hyperparameters to extract surfaces from an input \gls{mri} scan.
These methods work well and are thoroughly tested but have long run times due to their iterative nature and the need for topology correction. A representative example of the classical surface extraction process is FreeSurfer's widely used \texttt{recon-all} pipeline, which consists of several steps: first, a segmentation of the \gls{wm} is created and the boundary is tessellated to create an initial white matter surface, next the surface is corrected for topological errors, e.g., bridges or holes, to ensure the surface is continuously deformable to a sphere~\cite{fischl2001}, then the surface is locally deformed to match the intensity gradient of \gls{wm}-\gls{gm}-border visible in the input \gls{mri} scan, and finally the optimized \gls{wm} surface is deformed outwards to fit the \gls{gm} surface while monitoring and fixing self-intersections~\cite{dale1999}.


Recently, \gls{dl}-based models have emerged as an attractive alternative to the classical approaches, and are usually trained on their output. One of the main benefits of moving from an optimization-based approach to a learning-based approach is that the reconstruction time can be reduced from hours to seconds. The current \gls{dl}-based methods can be roughly divided into two main categories: implicit and explicit approaches. 

Implicit methods estimate a function that implicitly defines a surface. This function can be evaluated on the image grid and may reflect belief about a tissue class, e.g., \gls{wm}, in a certain voxel. This is the approach used in FastSurfer~\cite{henschel2020}, where the speed-up is obtained by predicting a volumetric segmentation using a \gls{cnn}. The segmentation is subsequently processed using steps similar to those in the classical approaches to obtain the surface representations. Instead of predicting a segmentation, it is also possible to predict a real-valued \gls{sdf} whose zero level-set defines a surface (e.g., the \gls{wm}-\gls{gm} interface) implicitly~\cite{gopinath2025}. This can be extended to arbitrary grid points, e.g., using hyper-columns~\cite{cruz2021}. Importantly, all implicit methods require a subsequent surface extraction step (e.g., using marching cubes)~\cite{lorensen1987}. The resulting surface, although accurately placed, is potentially subject to the same topological errors as the classical approaches and post-processing is typically necessary.

Explicit methods estimate a surface directly by learning to deform a template surface representation to fit the input image~\cite{wickramasinghe2020,bongratz2024,hoopes2022,santacruz2022}. They typically consist of two parts: a \gls{cnn} that extracts a set of feature maps from the input images, and a \gls{gcn} that estimates deformation vectors by sampling the feature maps at relevant locations. One of the main benefits of the explicit methods, as opposed to the implicit ones, is that the predicted surface is guaranteed to share its topology with the template surface. Consequently, if the template is topologically correct, so are the predicted surfaces, although geometrical errors such as \gls{sif} within and between surfaces can still occur. Additionally, explicit methods require that the template surface is co-registered to the input image in a preprocessing step. 

\paragraph{Extension to clinical data.}
Broadening the scope of surface reconstruction tools from \ToneWoneMM{} scans to clinical data is challenging because the acquisition parameters are not standardized in clinical examinations, which results in \gls{mri} scans with varying resolution and contrast. In other medical image analysis tasks, such as segmentation~\cite{billot2023}, registration~\cite{hoffmann2022}, and super-resolution~\cite{iglesias2023}, the capacity for \glspl{cnn} to generalize has been successfully expanded by using a \emph{domain randomization} approach where the training data is synthesized from label images and no real \gls{mri} scans are needed~\cite{gopinath2024a}. When the model is trained using synthetic data, the \glspl{cnn} generalize well to unseen data without the need for retraining. This approach is used in \cite{gopinath2025} where the authors present \gls{rac}, which is a variation of FreeSurfer's \texttt{recon-all} pipeline aimed at processing scans of arbitrary contrast and resolution. \gls{rac} predicts an \gls{sdf} for each cortical surface. The surface representation is then extracted from the \gls{sdf} using the same classical surface fitting steps as in \texttt{recon-all}. Because of the surface extraction steps, the average runtime of \gls{rac} is 3 to 4 hours on a standard desktop~\cite{gopinath2025}. As far as we are aware, \gls{rac} is the only tool that can currently do surface reconstructions from clinical \gls{mri} data.

A critical challenge when applying surface extraction tools to clinical data, or to data of varying quality in general, is that it can be difficult to assess accuracy of the estimated surfaces. The \gls{mri} data quality can vary both within a scan, e.g., when the scan has high in-plane resolution but large slice thickness, and across scans, e.g., due to motion artifacts. Consequently, having a direct measure of surface placement uncertainty would be extremely valuable, especially when processing large datasets where manual quality control becomes unfeasible. In \cite{gopinath2025unsurf}, the authors extend \gls{rac} with uncertainty estimation based on the difference between the \gls{sdf} estimated by the \gls{cnn} and the final \gls{sdf} computed from the extracted surface. Thus, the estimate does not directly reflect surface placement uncertainty due to the data (aleatoric uncertainty), but mostly highlights areas where extensive geometric processing was needed when extracting the surface from the initial \gls{sdf}. This approximation scheme is necessary to reduce computation, as directly estimating the surface placement uncertainty would require running the surface extraction step, which takes approximately 2 hours, multiple times from different \gls{sdf} samples. 

\paragraph{Contribution.} Here, we present a standalone tool for cortical surface reconstruction and spherical registration with integrated uncertainty estimation, which runs in less than \qty{1}{\second} on a GPU, and works on heterogeneous input scans ranging from low resolution clinical scans to high resolution research scans. This work extends our preliminary results presented at the International Workshop on Machine Learning in Medical Imaging~\cite{nielsen2026} in several aspects:
\begin{itemize}
    \item We present extensive comparisons against other state-of-the-art deep learning surface reconstruction tools on 15 different datasets consisting of high-resolution, research-quality \ToneWoneMM{} scans, and against \gls{rac} on low resolution clinical datasets. 
    \item We extend the model to predict an estimate of surface placement uncertainty, and we demonstrate that this correlates well with local prediction errors on \ToneWoneMM{} scans, and overall image quality on lower resolution clinical scans.
    \item We train the model to predict a spherical registration between the input scan and the \emph{fsaverage} surface template, and show that it achieves a registration accuracy that is similar to or better than state-of-the-art registration approaches on both \ToneWoneMM{} and lower resolution clinical data.
    \item We demonstrate the robustness of our approach in terms of contrast by evaluating the model on two well-known \emph{ex vivo} datasets; the BigBrain~\cite{amunts2013}, which is a 3D reconstruction of histological slices, and the Edlow brain~\cite{edlow2019}, which is an \gls{mri} scan.
\end{itemize}
As a reference for evaluating our method, we use surfaces reconstructed using FreeSurfer 7.4.1 applied to \ToneWoneMM{} scans which, in this context, can only be considered a ``silver standard''. We discuss the implications of this where relevant.

\section{Methods}
Our goal is to estimate cortical surfaces (\gls{wm} and \gls{gm}) from any \gls{mri} scan of the brain. Specifically, given an input image and a template surface we seek a model that deforms the vertices of the template surface so as to fit the input image. To simplify the theoretical description, we make a couple of assumptions. First, the input image should be oriented such that the linear part of its affine transformation (from voxel to world space) is a $3 \times 3$ identity matrix (i.e., \gls{ras} orientation, \qtyproduct[product-units = power]{1x1x1}{\milli\meter} voxels). Second, the template surface, which exists in \gls{mni} space, must be aligned with the input image beforehand. In practice, these two preprocessing steps are automatically handled by our method at inference time.

\subsection{Problem Formulation}
\label{sec:problem_formulation}
The \gls{wm} and \gls{gm} surfaces, which are 2D manifolds embedded in 3D Euclidean space, are defined by vertices and their connectivity. We use $\mathcal{M} = \left\{ \mathcal{V}, \mathcal{F} , \mathcal{E} \right\}$ to denote a surface mesh where $\mathcal{V} = \{v_1, v_2, \dots, v_n\}$, $\mathcal{F} = \{f_1, f_2, \dots, f_m\}$, $\mathcal{E} = \{e_1, e_2, \dots, e_q\}$ are the sets of vertices, faces, and edges, respectively. We use $\lvert \cdot \rvert$ to denote the cardinality of a set. $v_i \in \mathbb{R}^3$ are the coordinates of the $i$th vertex, $f_i \in \mathbb{N}^3$ is a tuple of vertex indices, and $e_i \in \mathbb{N}^2$ is a pair of vertex indices. Similarly, let $V \in \mathbb{R}^{n \times 3}$, $F \in \mathbb{N}^{m \times 3}$, and $E \in \mathbb{N}^{q \times 2}$ be matrices containing the vertices, faces, and edges of the manifold, respectively.

We assume that we are given the initial location of a set of vertices, $V_0$, which can be obtained from an affine registration of the template to the target scan, and we need to estimate, for each vertex, a (series of) velocity vector(s), $dV_t/dt$, which align the vertices with the anatomical boundaries in the input scan. This problem can be described by an \gls{ode} of the form~\cite{santacruz2022}
\begin{equation}
    \label{eq:surface-ode}
    \frac{d V_t}{dt} = g\left(s\left[ f(I), V_t \right]\right),
\end{equation}
where $I$ is the input \gls{mri} scan, $t \in \left[ 0, 1 \right]$ is the time, and $V_t$ are the vertex positions at time $t$. The function $f$ returns a number of features extracted from $I$, $s(A,B)$ is a function that samples a volume $A$ at coordinates $B$, and $g$ uses these (sampled) features to estimate the change in $V_t$. In this work, $f$ is a \gls{cnn} with a standard U-Net architecture, $g$ is a \gls{gcn}, and $s$ implements trilinear sampling (see \cref{fig:model_overview} and section \cref{sec:implementation} for details). At inference time, this is an initial value problem where the initial condition is the vertex positions of the template surface, i.e., $V_0 = V_\text{template}$. During training, predictions are supervised by the target surface by encouraging $V_1 \approx V_\text{target}$.

To solve \cref{eq:surface-ode}, i.e., to estimate $V_1$, we discretize the time interval in $K$ steps and apply a simple forward Euler integration scheme with step size $h=1/K$,
\begin{equation}
    \label{eq:forward-euler}
    V_{(k+1)h} = V_{kh} + h \times g(s\left[f(I), V_{kh}\right]).
\end{equation}
It is also possible to apply more advanced Runge-Kutta methods at the expense of increased computational load and memory consumption.

In addition to estimating the vertex positions as described above, we also wish to estimate two associated quantities, namely (1) the (aleatoric) uncertainty, $\sigma$, of the position of each vertex and (2) the spherical coordinates (registered to the \emph{fsaverage} template), $S$, of each vertex. To ensure that $\sigma$ is not negative, the network internally calculates $\log\sigma$ to which we apply the exponential function element-wise before returning it. Augmenting \cref{eq:surface-ode} with these quantities, we get
\begin{align}
    \begin{bmatrix} \dfrac{dV_t}{dt} \; \dfrac{d\log\sigma_t}{dt} \; \dfrac{dS_t}{dt} \end{bmatrix}^\top = g\left(s\left[ f(I), V_t \right]\right)
    \label{eq:surface-ode-general}
\end{align}
which we solve similarly to \cref{eq:forward-euler} except that we project the spherical coordinates back on the sphere after updating their positions. 

We set the initial condition of the uncertainty to $\log\sigma_0 = \begin{bmatrix} 0 \; 0 \; 0 \end{bmatrix}^\top$ (per vertex) corresponding to the uncertainty along each major axis. During training, as we have no target for this quantity, we cannot supervise its estimation directly so we optimize it based on the target vertex position (see \cref{sec:losses} for details). For the spherical coordinates, we take advantage of the fact that our template can be registered to the \emph{fsaverage} template thus associating a set of spherical coordinates with each vertex. We use these as $S_0$. To obtain $S_1$ during training, we use the spherical coordinates of the \emph{closest vertex on the target surface}. As described in \cref{sec:datasets}, all surfaces were resampled to the template. Consequently, these coordinates are the same for all subjects.



\subsection{Model Training}

\subsubsection{Domain Randomization}
\label{sec:data-synthesis}

We apply domain randomization as presented in~\cite{billot2023} and our previous work~\cite{nielsen2026}. Here we provide a brief description (see the supplementary material for more details).

We generate synthetic \gls{mr} images by sampling intensity values from a normal distribution conditioned on a label image. We apply smoothing with a Gaussian kernel to simulate partial volume effects and avoid sharp borders between regions with different labels. Finally, we sample a standard deviation for each label and use this to add normally distributed noise. We force a minimum contrast of 10 between \gls{wm}, \gls{gm}, and \gls{csf} to stabilize the training before applying a Gamma transformation and adding a slowly varying bias field.
Finally, we apply resolution augmentation to simulate different scanning protocols~\cite{iglesias2023}. Specifically, we randomly select one of the following scenarios: (1) no augmentation (\qty{1}{\milli\meter\cubed} voxels), (2) clinical scans (thick slices of \qtyrange{2.5}{8.5}{\milli\meter} with \qty{1}{\milli\meter\squared} in-plane resolution), (3) axial low field sequence (thick axial slices of \qtyrange{5}{5.4}{\milli\meter} \qty{1}{\milli\meter\squared} in-plane resolution), (4) near-isotropic low field sequence (\qtyrange{2}{5}{\milli\meter} along each axis). To provide a consistent input to the network, images are always resampled to \qty{1}{\milli\meter\cubed} voxels using trilinear interpolation, and intensities are normalized to values between 0 and 1. 
To increase robustness to inaccurate template registration, we perturb the template vertex coordinates by randomly drawn affine paramters.

\begin{figure*}
    \centering
    \includegraphics[width=\linewidth]{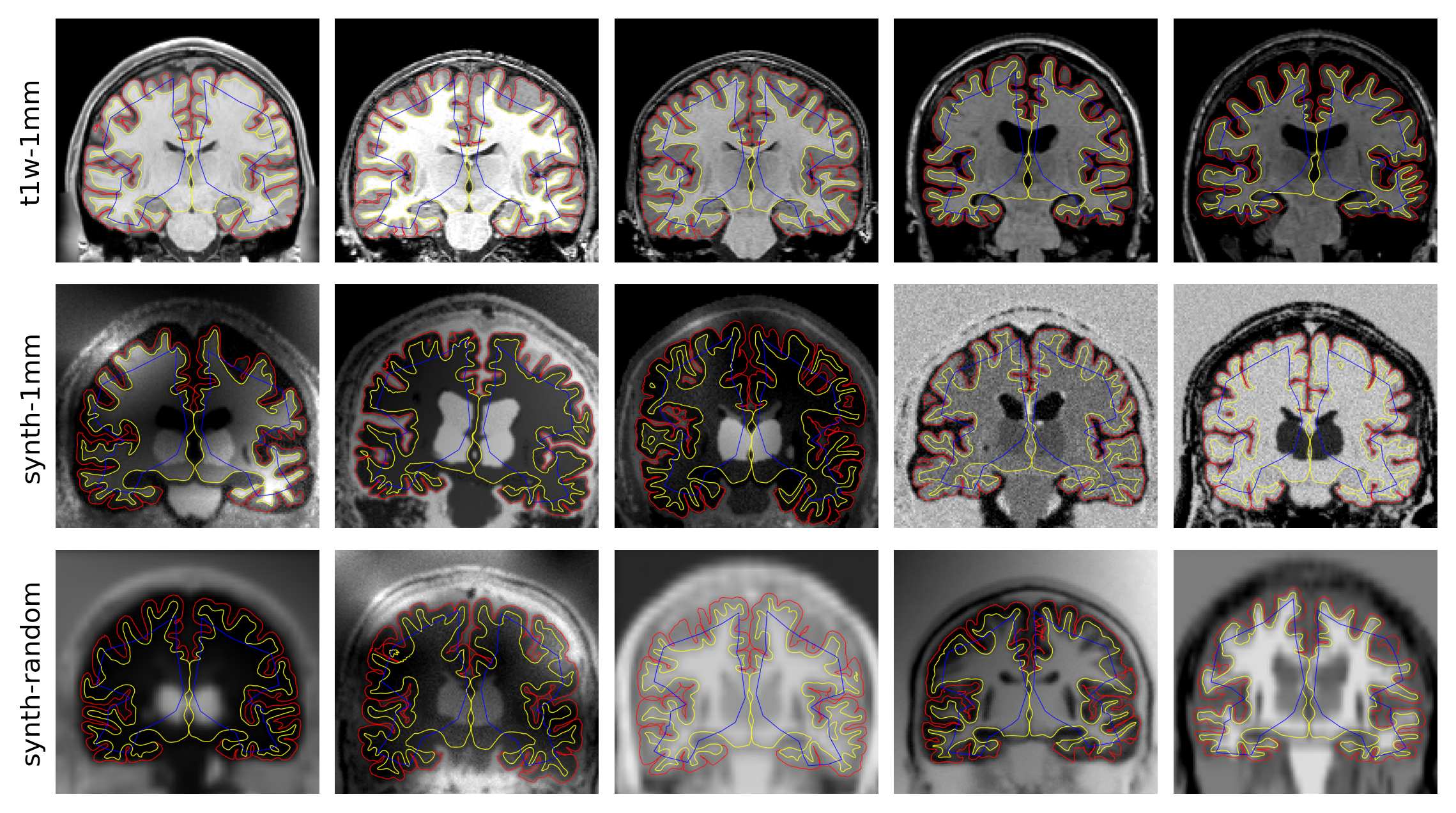}
    \caption{Examples of input data used to train the different models. The input template (blue) as well as the target \gls{wm} (yellow) and \gls{gm} (red) surfaces are overlaid on the input scans.}
    \label{fig:training_data_examples}
\end{figure*}

\Cref{tab:augmentation} gives an overview of the image augmentation steps used when training the different models and \cref{fig:training_data_examples} shows example outputs of these augmentation pipelines, i.e., the data used to train the different models.

\begin{table}
    \centering
    \begin{tabular}{p{1em} @{} l S[table-format=1.2] c c c}
        \toprule
        &&& t1w & synth & synth \\ 
        && $p$ & 1mm & 1mm & random \\ 
        \midrule
        \multicolumn{2}{l}{Select input image}          &       & x & x & x \\
        & T1w            &       & x &   &   \\
        & Labeling       &       &   & x & x \\
        \multicolumn{2}{l}{Sample intensities}    &       &   & x & x \\
        \multicolumn{2}{l}{Gamma transform}             & 0.33  &   & x & x \\
        \multicolumn{2}{l}{Extracerebral aug.}  & 0.20   & x & x & x \\
        & Skullstrip                                    & 0.50   & & & \\
        & Noise                                         & 0.50   & & & \\
        \multicolumn{2}{l}{Adjust FOV}                  &       & x & x & x \\ 
        \multicolumn{2}{l}{Bias field}                  & 0.75  &   & x & x \\
        \multicolumn{2}{l}{Resolution aug.}     & 0.75      &   &   & x \\
        & Clinical                                      & 0.33  & & & \\
        & Low field (axial)                             & 0.33  & & & \\
        & Low field (isotropic)                         & 0.33  & & & \\
        \multicolumn{2}{l}{Intensity normalization}     &       & x & x & x \\
        \bottomrule
    \end{tabular}
    \caption{Data augmentation steps for input scan when training the different models. FOV adjustment ensures that the data matrix is valid for processing with the \gls{cnn}. $p$ denotes the probability that an operation is applied (blank indicates that it is always applied). Indentation denotes a sub-step and x indicates that a step applies to this pipeline (if no sub-steps are marked, all sub-steps apply).}
    \label{tab:augmentation}
\end{table}

\subsubsection{Losses}
\label{sec:losses}
During training, we seek to find the network parameters that minimize a combination of data fidelity and regularization losses. We first describe the losses related to estimating the cortical surfaces and subsequently introduce the losses relevant to the spherical coordinate prediction.

\paragraph{Data fidelity losses.} To simplify notation, we define two functions, $\CPI$ and $\CP$. $\CPI$ returns the index of the point in a set of points, $\mathcal{P}$, that is closest to another point, $y$, in a Euclidean sense, i.e.,
\begin{equation}
    \CPI(\mathcal{P},y) = \underset{p \in \mathcal{P}}{\arg\min} \lVert y-p \rVert ^ 2,
    \label{eq:fun_closest_point_index}
\end{equation}
and $\CP$ returns the actual point
\begin{equation}
    \CP(\mathcal{P},y) = \mathcal{P}(\CPI(\mathcal{P},y)).
    \label{eq:fun_closest_point}
\end{equation}
Here $\lVert \cdot \rVert$ denotes the 2-norm. In the following, we use $\mathcal{P}_x$ ($\mathcal{P}_y$) to denote a set of 100,000 points sampled from the predicted (target) cortical surface. Points are sampled from each triangle with a probability proportional to its area. Similarly, $\mathcal{S}_x$ ($\mathcal{S}_y$) and $\mathcal{Z}_x$ ($\mathcal{Z}_y$) are the spherical coordinates and uncertainties corresponding to the points in $\mathcal{P}_x$ ($\mathcal{P}_y$) (obtained with barycentric interpolation). Note that we have no uncertainty on the target surface; therefore, $\mathcal{Z}_y$ is obtained by assigning the value of $\mathcal{Z}_x$ from the closest point in $\mathcal{P}_x$, i.e., $\mathcal{Z}_y = \left\{ \mathcal{Z}_x \left(\CPI\left[\mathcal{P}_x, p \right]\right) \mid p \in \mathcal{P}_y \right\}$.

In the initial training phase, we prefer to fit the vertex locations rather than their uncertainties. For this purpose, we use the standard symmetric chamfer distance,
\begin{equation}
    \label{eq:chamfer_eucl}
    \CD(\mathcal{P}_x, \mathcal{P}_y) = \dfrac{1}{\CARD{\mathcal{P}_x}} \sum_{p \in \mathcal{P}_x} \left\lVert p - \CP\left(\mathcal{P}_y, p\right) \right\rVert^2
\end{equation}
\begin{equation}
    \label{eq:chamfer_eucl_loss}
    L_\text{chamfer}= \CD(\mathcal{P}_x, \mathcal{P}_y) + \CD(\mathcal{P}_y, \mathcal{P}_x),
\end{equation}
with a Euclidean distance function.

Next, to estimate the data related (aleatoric) uncertainty associated with the predicted vertex coordinates, we follow \cite{hu2025hierarchical} and make the assumption that the error is normally distributed and independent over vertices in each of the three spatial dimensions. Therefore, we need to estimate, for each vertex, a mean (location) and a standard deviation for each dimension (uncertainty of location). To do this, we use the negative log-likelihood as our distance function,
\begin{multline}
    -\mathcal{L}(\mu, \Sigma \mid x) = -\log p(x \mid \mu, \Sigma) = \\
    \dfrac{1}{2} \left( d \log \left( 2\pi \right) + \log\det\Sigma + \left(x-\mu\right)^\top \Sigma^{-1}\left(x-\mu\right) \right).
\label{eq:nll}
\end{multline}
Here $d=3$ and, since we assume independent errors, $\Sigma$ is a diagonal matrix which simplifies the evaluation of this expression considerably. Note that by setting $\Sigma=I$ (where $I$ is the identity matrix) and dropping the constant terms, this reduces to a quantity which is proportional to the squared Euclidean distance. We construct a loss function from $-\mathcal{L}$ by defining
\begin{multline}    
    \CD\left( \mathcal{P}_x, \mathcal{P}_y, \mathcal{Z} \right) = \\
    \dfrac{1}{\CARD{\mathcal{P}_x}} \sum_{p \in \mathcal{P}_x, \Sigma \in \mathcal{Z}_x} -\mathcal{L}\left(
        p, \Sigma \mid \CP\left( \mathcal{P}_y, p \right) \right)
    \label{eq:chamfer_nll}
\end{multline}
\begin{equation}
    L_{-\mathcal{L}} = \CD\left(\mathcal{P}_x, \mathcal{P}_y, \mathcal{Z}_x\right) + \CD\left(\mathcal{P}_y, \mathcal{P}_x, \mathcal{Z}_y\right)
    \label{eq:chamfer_nll_loss}
\end{equation}
where $\mathcal{Z}_x$ ($\mathcal{Z}_y$) is a set of $\Sigma$ associated with each point in $\mathcal{P}_x$ ($\mathcal{P}_y$).
Whereas \cref{eq:chamfer_eucl} uses the classic Chamfer distance, \cref{eq:chamfer_nll} is different because we find the closest point using the Euclidean distance but compute the distance using $-\mathcal{L}$.

\paragraph{Regularization losses.} We use three regularization terms to encourage smoothness, even distribution of vertices, and high quality triangles on the predicted surfaces.


To encourage smoothness, we use a loss that penalizes the mean squared Euclidean distance between a predicted surface and its Taubin smoothed counterpart. Taubin smoothing \cite{taubin1995} acts as a low-pass filter without introducing shrinkage (as opposed to, for example, Laplacian smoothing),
\begin{equation}
    V^N = \left( \left( I - \mu K \right) \left( I - \lambda K \right) \right)^N V,
\end{equation}
where $K = I - W$ is an $n \times n$ matrix and $W$ is a weight matrix. Here, we use the adjacency matrix of the graph as $W$. The free parameters define the properties of the filter and we use $\lambda=0.33$, $\mu=-0.34$, and $N=10$ where $N$ is the number of iterations. The loss is therefore
\begin{equation}
    L_{Taubin} = \dfrac{1}{n}\sum_{i=1}^{n} \lVert v_i - v_i^N\rVert^2,
\end{equation}
where $v_i$ and $v_i^N$ are corresponding vertices on the surface before and after Taubin smoothing, respectively. This loss promotes smoothness without (implicitly) encouraging a particular shape. This is different from, for example, the normal consistency loss (e.g. \cite{bongratz2024, hoopes2022}) which penalizes any curvature deviation from zero.

We use an edge loss to encourage equally sized triangles by penalizing the variance of the normalized edge lengths. Let $e = (i,j)$ be the vertex indices of the edge $e$, then we define the edge length and the normalized edge length of edge $e$ as follows
\begin{align}
    l_e &= \lVert v_i - v_j \rVert \\
    \bar{l}_e &= q\frac{l_e}{\sum_{e^\prime \in \mathcal{E}} l_{e^\prime}}
\end{align}
such that the mean over all $\bar{l}_{ij}$ is one. The corresponding loss is
\begin{equation}
    \label{eq:edge_length}
    L_\text{edge} = \dfrac{1}{q}\sum_{e \in \mathcal{E}} \left( 1 - \bar{l}_e \right)^2.
\end{equation}
The normalization is done to avoid shrinking the mesh, which may happen if the mean squared edge lengths are penalized directly.

To encourage high-quality (equilateral) triangles, we use the following metric~\cite{bhatia1990},
\begin{equation}
    \label{eq:tri_quality}
    \text{TQ}_i = 4 \sqrt{3} \frac{A_i}{\sum_{e \in \mathcal{E}(f_i)} l_e^2},
\end{equation}
where $A_i$ and $\mathcal{E}(f_i)$ denote the area and the set of edges belonging to the $i$th face, respectively. $\text{TQ}$ is 1 for an equilateral triangle. The corresponding loss we define as
\begin{equation}
    L_\text{quality} = 1 - \frac{1}{m}\sum_{i=1}^m \text{TQ}_i,
\end{equation}

Finally, we include a loss that encourages meaningful vertex correspondence between the white and gray matter surfaces. Let $d_i = v(\text{GM})_{i} - v(\text{WM})_i$ denote the vector pointing from a vertex on the \gls{wm} surface to the corresponding vertex on the \gls{gm} surface and let $n_i$ be the normal of $v(\text{WM})_i$. We penalize the cosine of the angle between these vectors in the following way, 
\begin{equation}
    L_\text{cos} = \dfrac{1}{n}\sum_{i=1}^{n} \left( 1 - \frac{n_i^\top d_i}{\left\lVert n_i \right\rVert \left\lVert d_i \right\rVert } \right)^2.
\end{equation}
This loss serves three purposes. First, it affords us some control over the vertex-to-vertex correspondence between the cortical surfaces. When coupled with the smoothness encouraged by $L_\text{Taubin}$, it is essentially a simplified version of the procedure proposed in \cite{fischl2018} to establish meaningful white-gray matter correspondence.
Second, it discourages intersections between the \gls{wm} and \gls{gm} surfaces (provided that the curvature is not extremely high). Finally, and related to this, it encourages decoupling of the cortical surfaces (e.g., at the midline where it pushes the \gls{gm} vertices slightly outwards), thus providing an interior domain that is suitable for solving partial differential equations, e.g., using the finite element method where the interior domain needs to be meshed.


\paragraph{Spherical registration losses.} To learn the spherical coordinates associated with each vertex, we use three different losses. The data fidelity term is a modified version of the Chamfer distance where we find closest points on the \gls{wm} surfaces but compute the distance between the spherical coordinates of these points. Let $\mathcal{W} = \left\{ 1 / \lVert \text{diag}(\Sigma) \rVert \mid \Sigma \in \mathcal{Z} \right\}$ be the set of inverse norm of the uncertainties, then we use the following spherical chamfer loss
\begin{multline}
    \CD\left(\mathcal{P}_x, \mathcal{P}_y, \mathcal{S}_x, \mathcal{S}_y, \mathcal{W}\right) = \\
    \dfrac{1}{\sum_{w \in \mathcal{W}} w} \sum_{p \in \mathcal{P}_x, s \in \mathcal{S}_x, w \in \mathcal{W}} w \left\lVert s - \mathcal{S}_y\left(\CPI\left[\mathcal{P}_y, p\right]\right) \right\rVert^2
\end{multline}
\begin{equation}
    L_\text{sphere} = \CD\left(\mathcal{P}_x, \mathcal{P}_y, \mathcal{S}_x, \mathcal{S}_y, \mathcal{W}_x\right) + \CD\left(\mathcal{P}_y, \mathcal{P}_x, \mathcal{S}_y, \mathcal{S}_x, \mathcal{W}_y\right) 
\end{equation}
Inspired by FreeSurfer's registration approach \cite{fischl1999a}, we include two regularization terms. The first term penalizes metric distortions by encouraging similar edge lengths between the \gls{wm} surface and its spherical representation,
\begin{align}
    s &= \dfrac{\sum_i e_i}{\sum_i e_i^0} \\
    L_\text{distortion} &= \dfrac{1}{\lvert E \rvert}\sum_{i=1}^{\lvert E \rvert} \left( e_i - s \cdot e_i^0 \right)^2.
\end{align}
Here, $e_i$ is the length of the $i$th edge on the predicted sphere, $e_i^0$ is the length of the $i$th edge on the predicted \gls{wm} surface, and $s$ scales the sum of $e^0$ to that of $e$.

The second term discourages folded triangles on the sphere by penalizing area differences between triangles on the \gls{wm} surface, $A^0$, and on the sphere, $A$, i.e.,
\begin{equation}
    L_\text{area} = \sum_{i \in F} p(A_i)\left( A_i - A_i^0 \right)^2; \quad p(A) =
        \begin{cases}
            1 & \text{if } A < 0 \\
            0 & \text{otherwise}.
        \end{cases}
\end{equation}
Here, $A_i$ is the \emph{oriented} area of the $i$th face on the sphere. It is negative if its normal points towards the center of the sphere.

\subsubsection{Implementation}
\label{sec:implementation}
Following~\cite{hoopes2022} (with minor modifications), we model $f$ in \cref{eq:surface-ode} using a \gls{cnn} with a UNet architecture~\cite{ronneberger2015}. The UNet has five levels (encoder channels 16, 32, 64, 128, 256; decoder channels 128, 64, 32, 16)
between which we do max-pooling with a kernel size of 2 in the encoder and upsampling with a factor of 2 in the decoder. At each level, $3 \times 3 \times 3$ convolutions are applied followed by instance normalization and \gls{prelu} activation. We use the feature maps from all levels of the decoder.

To estimate the \gls{wm} surface, we perform integration of the quantities of interest (vertex position, vertex position uncertainty, and spherical coordinates) using a multiscale approach where we iteratively update and upsample these quantities. The initial template has 62 vertices (scale $n=1$) and the final surface prediction has 245,762 vertices (scale $n=7$). To upsample from scale $n$ to $n+1$, we insert a new vertex at the midpoint of each edge, thus splitting each face into four new faces. Because of this multiscale approach, we have not only a single function $g$ in \cref{eq:surface-ode-general} but one for each scale, $g_n$, all of which are modeled with a \gls{gcn} with a UNet-like architecture which has a maximum depth of four levels (encoder channels 96, 96, 96, 96, max-pooling; decoder channels 96, 96, 96, max-unpooling). For example, when $n=1$ (coarsest scale) there is only one level ($\left\{1\right\}$) and when $n=7$ (finest scale) there are four levels ($\left\{7,6,5,4,5,6,7\right\}$). The graph convolutions are implemented on the edges to immediate neighbors \cite{wang2019} followed by instance normalization and \gls{prelu} activation. Finally, a graph convolution layer is used to estimate the deformation vectors. To summarize, we divide the interval $t \in \left[ 0, 1 \right]$ into $N=7$ pieces, each of which is solved in $K=2$ steps as described in \cref{eq:forward-euler} (this could be unrolled by setting $K=NK$). This is presented as an algorithm in the supplementary material.

\begin{figure*}
    \centering
    \includegraphics[width=\linewidth]{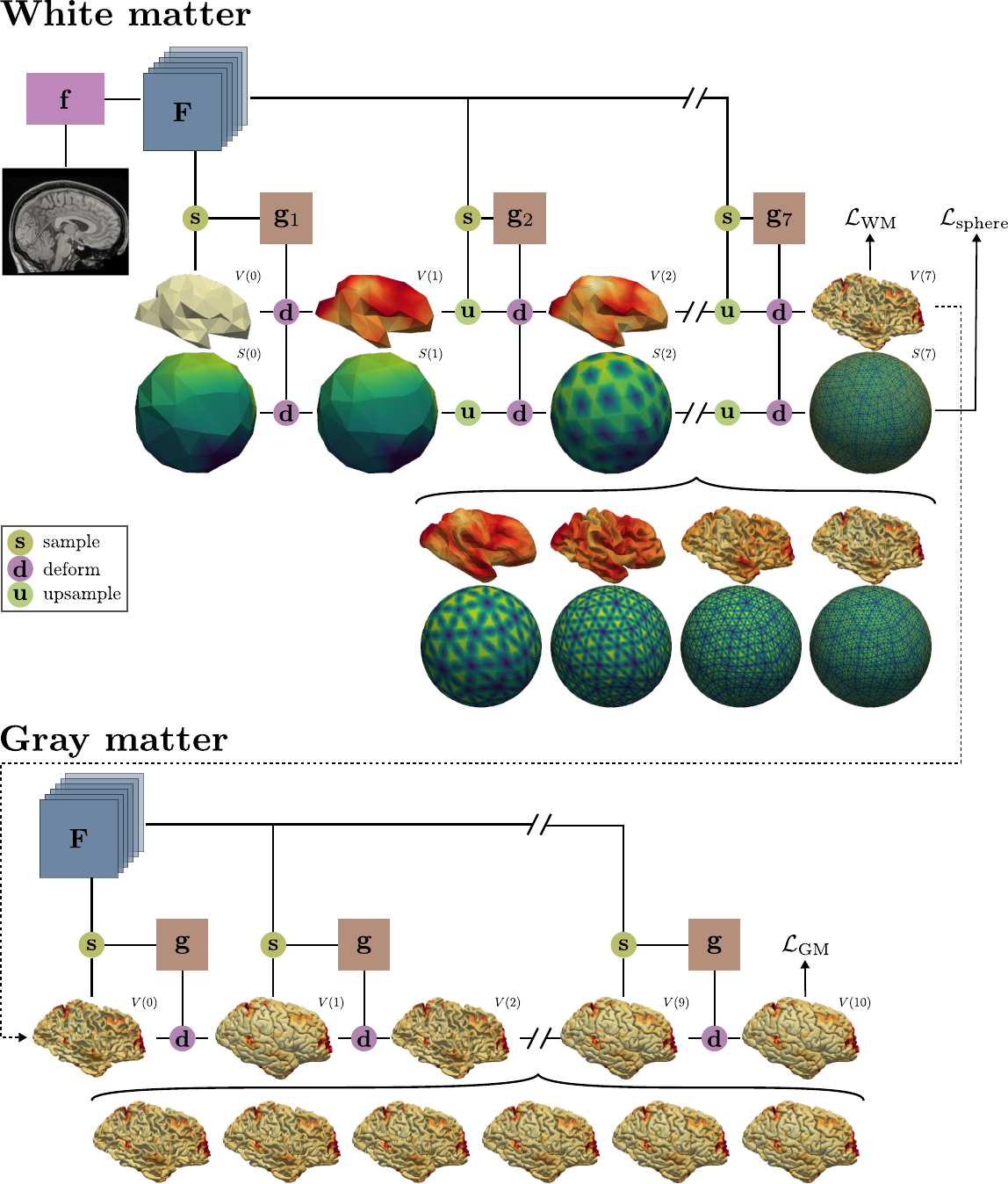}
    \caption{Overview of network training and inference. The input scan (here we show a single slice of a \gls{t1w} image) is just an example; for the synth models this could be any contrast and for the synth-random model the resolution could be different. $\mathbf{f}$ is the image feature extractor (a UNet in our case) which extracts the features $\mathbf{F}$. $g_i$ are the \gls{gcn} blocks which predict the vectors used to deform the input template surface. Note that $V(7)$ from the \gls{wm} pipeline is the final \gls{wm} surface prediction and serves as the input for the \gls{gm} pipeline. The predicted uncertainty is used to color the cortical surfaces as they evolve; the sphere is simply colored by vertices index to show its structure (we do not color by, for example, convexity or curvature since this is not used during training). $\mathcal{L}_x$ refers to the losses described in \cref{tab:loss_weights} (the \gls{wm}-\gls{gm} loss is not shown).}
    \label{fig:model_overview}
\end{figure*}

The \gls{gm} surface is then estimated from the \gls{wm} surface using only a single $g$ since this displacement is relatively simple compared to predicting the \gls{wm} surface from the template positions. In this case, $g$ is a linear layer with 32 channels followed by \gls{prelu} activation and another linear layer to predict the displacement vectors. We repeat this process $K=10$ times. This is presented as an algorithm in the supplementary material and the complete model is shown in \cref{fig:model_overview}.


Since the squared difference between predicted and target coordinates change by a few orders of magnitude during training, we gradually change the weighting of the losses as training progresses. We start by training on the (standard) Chamfer loss since at this stage the uncertainties would be high simply because we are far from ``convergence''. After a while, we switch to the $-\mathcal{L}$-based distance loss and when the uncertainties have achieved sensible values, we use these to weigh the spherical coordinate Chamfer distance. An overview of the loss weights and scheduling can be found in \cref{tab:loss_weights}.
We set slightly lower weights of $L_\text{edge}$ for the \gls{gm} surface compared to the \gls{wm} surface to allow for better vertex-to-vertex correspondence.

The training is done on an Nvidia RTX 6000 GPU (\qty{48}{\giga\byte} RAM) using the AdamW optimizer ($\lambda = 0.01$; $\beta_1=0.9$; $\beta_2=0.999$) with an initial learning rate of $10^{-4}$ which we decrease to $5\times10^{-5}$ after 20,000 gradient steps and further to $2.5\times10^{-5}$ after 60,000 gradient steps. We use a batch size of 1 and train until the standard chamfer distance on the validation set plateaus which was approximately 80,000 gradient steps ($\approx 24$ hours). The model is implemented in PyTorch 2~\cite{ansel2024} and uses the PyTorch-Ignite framework~\cite{fomin2020} for training. To speed up model convergence, we start by training a model from scratch on \gls{t1w} scans. The image feature extractor ($f$ in \cref{eq:surface-ode}) of the model using synthetic data is then pretrained using a mean squared error loss of the feature maps (comparing the features from the to-be-trained network on a synthetic scan and the features from the pretrained network on a corresponding \gls{t1w} image). This is fast ($\approx 6$ hours) and, in our experience, makes training more stable, especially when training on low-resolution, synthetic data. Once trained, the inference time is approximately 0.6 seconds on a GPU.

At inference time, we need to register the template mesh (which exists in \gls{mni} space) to the input \gls{mri} scan. Therefore, inspired by the DeepSurfer package \cite{hoopes2022}, we train a small network that performs an affine registration of a brain \gls{mri} scan to \gls{mni} standard space. This network is trained on synthetic data with resolution augmentation and can be run before the main surface prediction network to align the template to the input scan.

\begin{table}
\caption{Overview of the loss weights at each stage of the training process. The first number in the ``weight'' columns denotes the gradient step after which those weights apply. (w) indicates that it is weighted by the uncertainty estimate.}
\centering
\begin{tabular}{l l S S S}
    \toprule
    \textbf{Task} & \textbf{Loss} & \multicolumn{3}{c}{\textbf{Weight}} \\
    &  & {0} & {20000} & {40000} \\
    \midrule
    WM      & chamfer & 1.0   & 0.0   & 0.0 \\
            & $-\mathcal{L}$     & 0.0   & 0.5   & 0.5 \\
            & Taubin  & 40.0  & 40.0  & 40.0 \\
            & edge    & 5.0   & 2.0   & 1.0 \\
            & quality & 2.5   & 1.0   & 0.5 \\
    GM      & chamfer & 1.0   & 0.0   & 0.0 \\
            & $-\mathcal{L}$  & 0.0   & 0.5   & 0.5 \\
            & Taubin  & 20.0  & 20.0  & 20.0 \\
            & edge    & 2.5   & 1.0   & 1.0 \\
            & quality & 2.5   & 1.0   & 0.5 \\    
    WM-GM & cos     & 1.0   & 1.0   & 1.0 \\
    Sphere  & chamfer       & 0.01 & 0.01 & 0.0 \\
            & chamfer (w) & 0.0 & 0.0 & 0.01 \\ 
            & distortion    & 0.01 & 0.01 & 0.01 \\
            & area          & 0.001 & 0.001 & 0.001 \\
    \bottomrule
\end{tabular}
\label{tab:loss_weights}
\end{table}

\subsection{Experiments}

\subsubsection{Datasets}
\label{sec:datasets}
We train our model on multiple datasets (see \cref{tab:train-datasets}) all of which include high-resolution (near \qty{1}{\milli\meter} isotropic) \gls{t1w} scans. All scans were resampled to \qtyproduct[product-units = power]{1x1x1}{\milli\meter} and processed with FreeSurfer 7.4.1 to obtain the \gls{wm} and \gls{gm} surfaces and a brain segmentation. To generate synthetic \gls{mri} scans (see \cref{sec:data-synthesis}), we simplify the volume segmentation by merging structures with similar intensity profiles (e.g., amygdala, cortex, hippocampus), and model the extracerebral tissues using $k$-means clustering. The cortical surfaces, which are the training targets, were resampled to the template surface, which the model learns to deform. For training, we coregister the template to the input scan by calculating an affine transformation that minimizes the squared distance between the template and the surface of that particular subject. Since all surfaces were resampled to the template, we solve for the $4 \times 4$ matrix which, when applied to the template points, minimizes the difference between the template and the target points (augmented with a column of ones) in a least squares sense. The training dataset is split in train, test, and validation subsets with fractions 0.8, 0.1, and 0.1. All datasets were subjected to manual quality control which led to some subjects being excluded (see \cref{tab:train-datasets}).

\begin{table}[t]
\centering
\begin{threeparttable}
    \caption{Overview of training and test datasets. Datasets in the ``train'' section were used for training, validation, and testing (splits 0.8, 0.1, 0.1). Datasets in the ``generalization'' section were used to test on unseen subjects from unseen datasets (generalizability). The number in the ``Diag.'' (diagnosis) column refer to the number of subjects with \emph{any} kind of diagnosis. ``Scans'' refers to the number of \emph{scans} that passed quality control.}
    \begin{tabular}{l S[table-format=4]@{\quad} S[table-format=4]@{\quad} S[table-format=4]@{\quad} l S[table-format=4]@{\quad}}
        \toprule 
        \textbf{Dataset} & \textbf{Subj.} & \textbf{Males} & \textbf{Diag.} & \textbf{Age} & \textbf{Scans} \\
        \midrule
        \multicolumn{5}{l}{\textbf{Train}} \\
        \footnotesize \acrshort{abide}~\cite{dimartino2014}   & 1112          & 1001  & 539  & $7-64$  & 807 \\
        \footnotesize ADHD200~\cite{milham2012}               & 947           & 575  & 362   & $7-26$  & 820  \\
        \footnotesize ADNI3~\cite{weiner2017}                 & 331           & 170  & 199   & $61-97$ & 284 \\
        \footnotesize \acrshort{aibl}~\cite{ellis2009}        & 668           & 294  & 668   & $55-96$ & 627  \\
        \footnotesize Buckner40~\cite{FISCHL2002341}                               & 40            & 19   & 0     & $18-93$ & 38  \\
        \footnotesize \acrshort{chinese-hcp}~\cite{vogt2023}  & 213           & 58   & 0     & $18-30$ & 212 \\
        \footnotesize \acrshort{hcp}~\cite{vanessen2013}      & 897           & 448  & 0     & $22-35$ & 821 \\
        \footnotesize \acrshort{oasis3}~\cite{lamontagne2019} & 1378\tnote{1} & 507  & 622   & $42-97$ & 1210  \\
        \midrule
        \multicolumn{5}{l}{\textbf{Generalization}} \\
        \footnotesize \acrshort{cobre}~\cite{mayer2013}       & 191\tnote{1}   & 110   & 72    & $18-65$ & 187 \\
        \footnotesize \acrshort{isbi2015}~\cite{carass2017}   & 5     & 1     & 5    & $44 \pm 10$ & 21\tnote{2} \\
        \footnotesize \acrshort{mcic}~\cite{gollub2013}       & 200   & 104   & 110  & $18-60$ & 161 \\
        \footnotesize \acrshort{aomic}~\cite{snoek2020}       & 928   & 445   &  0    & $19-26$ & 928 \\
        \midrule
        \multicolumn{5}{l}{\textbf{Generalization} (\qtyproduct[product-units = power]{1x1x5}{\milli\meter} \acrshort{flair})} \\
        \footnotesize \acrshort{adni}-GO2                     & 199   & 94    & 99   & $56-89$ & 199 \\
        \midrule
        \multicolumn{5}{l}{\textbf{Ex vivo}} \\
        \footnotesize BigBrain~\cite{amunts2013}                  & 1 & 1 & 0 & $65$ & 1 \\
        \footnotesize Edlow~\cite{edlow2019}                      & 1 & 0 & 0 & $58$ & 1 \\
        \bottomrule
    \end{tabular}
    \begin{tablenotes}
        \item [1] \footnotesize Demographic data only available for a subset of the subjects.
        \item [2] \footnotesize Some subjects contributed multiple scans.
    \end{tablenotes}
    \label{tab:train-datasets}
\end{threeparttable}
\end{table}

To quantify the performance between models trained on synthetic and real data, we train three different models. The first model, denoted \gls{t1w}-1mm, is trained on \ToneWoneMM{} scans; the second model, denoted synth-1mm, is trained on \qtyproduct[product-units = power]{1x1x1}{\milli\meter} synthetic scans; and the third model, denoted synth-random, is trained on synthetic scans of random resolution.

\subsection{Model Evaluation}
\subsubsection{Surface Accuracy and Quality}
We evaluate the accuracy of the predicted cortical surfaces by computing the \gls{assd}
\begin{multline}
    \text{ASSD}(\mathcal{M}_A,\mathcal{M}_B) = \\
    \frac{1}{\left| \mathcal{V}_A \right| + \left| \mathcal{V}_B \right|} \left( \sum_{v \in \mathcal{V}_A} d(v,\mathcal{M}_B) + \sum_{v \in \mathcal{V}_B} d(v,\mathcal{M}_A) \right)     
\end{multline}
and \gls{hd90}
\begin{multline}
    \text{HD90}(\mathcal{M}_A,\mathcal{M}_B) =
    \max \big( P_{90}\left\{ d(v,\mathcal{M}_B) | v \in \mathcal{V}_A \right\},  \\
    P_{90}\left\{d(v,\mathcal{M}_A) | v \in \mathcal{V}_B\right\} \big)     
\end{multline}
where $\mathcal{M}_A$ denotes a (surface) mesh $A$ with vertices $\mathcal{V}_A$, and $d(v,\mathcal{M})$ computes the distance from point $v$ to the closest point on $\mathcal{M}$ (not necessarily a vertex of $\mathcal{M}$). As reference, we use surfaces created with FreeSurfer's \texttt{recon-all} pipeline on \qtyproduct[product-units = power]{1x1x1}{\milli\meter} \gls{t1w} scans.

We also report several metrics reflecting different aspects of the quality of the surface meshes: the number of \gls{sif}, the number of intersections between \gls{wm} and \gls{gm} surfaces, triangle quality (calculated from \cref{eq:tri_quality}) and triangle shape (calculated from \cref{eq:edge_length}). We restrict the calculation of these quantities to valid surface areas, i.e., we ignore the medial wall.

We evaluate the models on the test set (unseen subjects from datasets used during training) and generalization set (unseen datasets).
We also report these numbers for a representative set of state-of-the-art \gls{dl} methods trained on \ToneWoneMM{} data, specifically, the TopoFit implementation in the \emph{DeepSurfer} package (version 0.1.1a5), V2C-Flow\footnote{\url{https://github.com/ai-med/Vox2Cortex} at commit \texttt{809c937}.} \cite{bongratz2024}, and \gls{rac}, a method specifically aimed at surface reconstructions of low-resolution data~\cite{gopinath2025}. DeepSurfer was trained on \gls{ixi}, \gls{mcic}, and OASIS-1 (788 subjects); V2C-Flow was trained on ADNI (1,155 subjects) \cite{bongratz2024}; \gls{rac} was trained on 1000 randomly selected subjects from \gls{adni} and \gls{hcp} (500 from each).

\subsubsection{Surface Uncertainty}
To evaluate the uncertainty estimates from our model, we correlate them with the surface reconstruction errors that we observe when comparing with the surfaces from FreeSurfer's \texttt{recon-all} pipeline. Note that although our model provides an uncertainty estimate along each axis of Euclidean space ($x$, $y$, and $z$) (see \cref{eq:nll}), here we always use the $l_2$ norm over these values, and this is what is referred to by ``uncertainty'' in the following.

\begin{figure}
\centering
\includegraphics[width=\linewidth]{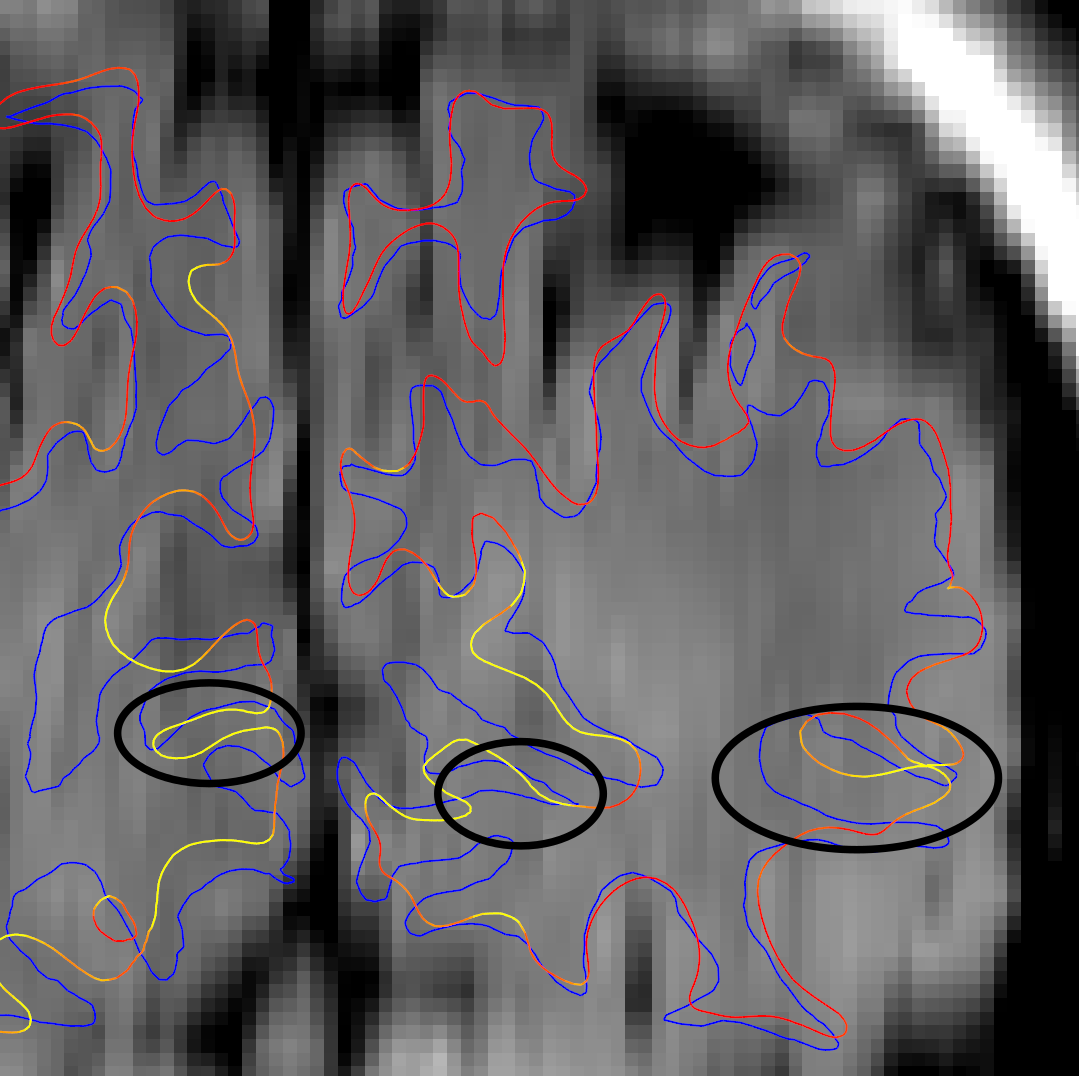}
\caption{Illustration of the problem of correlating closest distance with uncertainty when the surface reconstruction error is large. The blue surface is reconstructed using FreeSurfer on a corresponding \ToneWoneMM{} scan (not shown) whereas the red and yellow surface is estimated from a \FLAIRfiveMM{} scans (shown). Red and yellow indicate low and high uncertainty, respectively. The three encircled parts show examples of patches where the (closest) distance to the FreeSurfer surface is small even though the surface estimate is not correct thus warranting a high uncertainty. Spherical registration aligns the folding patterns of the surfaces and by resampling one surface onto the other, we hope to achieve a more appropriate alignment of the surfaces from which to calculate the distances.}
\label{fig:sphere-reg-distance-error}
\end{figure}

To estimate surface reconstruction errors of models applied to scans with a resolution of \qtyproduct[product-units = power]{1x1x1}{\milli\meter}, we use the distance from a vertex on the predicted surface to the closest point on the reference surface.
To estimate surface placement errors of models applied to scans of low resolution (specifically, the \FLAIRfiveMM{} scans from \gls{adni} GO-2), we map both predicted and target surfaces to the sphere, register the target surface to the predicted surface, resample the latter to the former, and finally compute distances between the corresponding vertices\footnote{Consequently, the actual values of the errors are higher than what is typically reported using \gls{assd}. However, what is interesting in this regard is not the actual value of the error but its relationship with the uncertainty predictions.}. The reason for doing this is that when the errors are large, the closest point on the reference surface may not be meaningful in the sense that this is not the point where that vertex was ``supposed'' to be (e.g., in the wrong sulcus). This is particularly problematic when correlating with uncertainty and is illustrated in \cref{fig:sphere-reg-distance-error}. 

To investigate if the uncertainty estimate can inform us about the general quality of a surface reconstruction for a given subject, we compare mean values of uncertainty and error (over the whole cortex) between subjects.
To investigate if the uncertainty estimate is useful for \emph{localizing} areas of poor surface accuracy within a subject, we correlate uncertainty estimates with errors on a per-vertex level. However, since we are interested in localizing patches of poor reconstruction (e.g., a gyral crown) rather than individual vertices \emph{per se}, we apply five iterations of Laplacian smoothing to the error estimate before calculating the correlation. The uncertainty estimates are generally smooth whereas the surface errors are not because they are sensitive \emph{any} deviation between the surfaces (some of which may be due to the exact location of the vertices). We always ignore the medial wall when reporting the metrics above.

Finally, we map uncertainties and errors to \emph{fsaverage} to visually compare their correspondence at the group level.

\subsubsection{Surface Registration}
To evaluate the spherical registration produced by our method, we use it to map the atlas from \cite{desikan2006} to the subject surface and calculate the fraction of correctly labeled vertices (``classification accuracy'') using a registration created with the default spherical registration approach in FreeSurfer as reference \cite{fischl1999a}. For comparison, we also compute the spherical registrations using the method from the \gls{cat12} toolbox \cite{yotter2011} and JOSA \cite{li2024}.
Additionally, since we deform a template mesh that has a predefined registration to \emph{fsaverage}, we can simply use these coordinates directly to obtain a spherical registration. We denote this as ``ours (map)''. As explained in \cref{sec:problem_formulation}, this is the starting point of our method; hence, the difference between ``ours (map)'', $S_0$, and final coordinate predictions by our method (called ``ours (pred)'', $S_1$) is what we learn.




\section{Results}
\begin{figure*}
    \centering
    \includegraphics[width=0.75\linewidth]{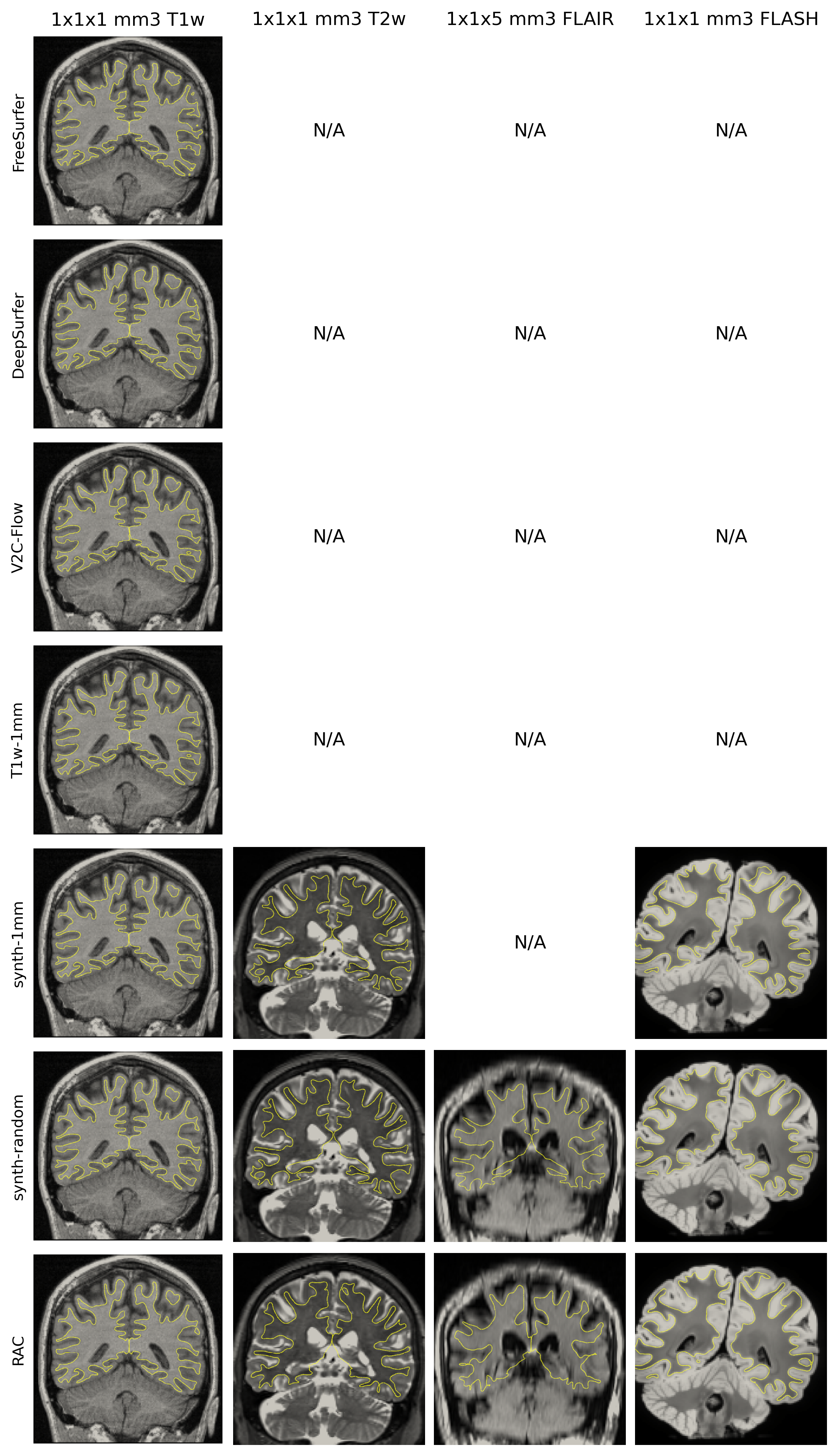}
    \caption{Comparison of different surface reconstruction methods showing which cases are handled by the different methods. The \emph{ex vivo} \gls{flash} scan is from \cite{edlow2019}. Its original resolution is \qtyproduct[product-units = power]{0.1x0.1x0.1}{\milli\meter} but downsampled to \qtyproduct[product-units = power]{1x1x1}{\milli\meter}. N/A denotes that the model is not applicable to the given contrast-resolution combination.}
    \label{fig:matrix_examples}
\end{figure*}

\Cref{fig:matrix_examples} illustrates which cases (e.g., contrast and resolution) are handled by the different models by showing examples of surface reconstructions. It shows predictions on \emph{in vivo} \gls{t1w} and \gls{flair} scans as well as an \emph{ex vivo} \gls{flash} scan.

\subsection{Surface Accuracy and Quality}
In \cref{tab:eval_accuracy}, we quantify surface accuracy and quality of our method, DeepSurfer, and V2C-Flow. Unsurprisingly, we see that the model trained specifically on \ToneWoneMM{} scans performs better than the models trained on synthetic scans when applied to this kind of data. In terms of mesh quality, our models produce surfaces with triangles of more equal size and shape compared to DeepSurfer and V2C-Flow. The surfaces from our method and DeepSurfer contain very few \gls{sif} and \gls{wm}-\gls{gm} intersections.

We also show results of applying our synth-random model to \FLAIRfiveMM{} from the \gls{adni} GO-2 dataset. As expected, the surface accuracy decreases compared to predicting on \ToneWoneMM{} scans, although the error is still below the image resolution ($\sim$\qty{0.70}{\milli\meter}) and lower than that of \gls{rac} ($\sim$\qty{0.94}{\milli\meter}). The mesh quality metrics, however, are comparable to those obtained on \ToneWoneMM{} data.

\Cref{tab:eval_accuracy} suggests that overall generalization performance of our model on unseen \ToneWoneMM{} scans is similar to that of the test set. However, it may still be interesting to explore performance on individual datasets to investigate whether the model is biased towards certain image characteristics, demographics, or scanning hardware. Hence, in the supplementary material we show \gls{assd} per dataset for the test and generalization sets. Performance is stable across datasets albeit slightly lower for synth-1mm and synth-random models on \gls{isbi2015} which is a small dataset containing lesions where the FreeSurfer-based ground truth is not particularly accurate (see \cref{sec:discussion} for more details).

Lastly, we show the results of applying our models as well as \texttt{RAC} to two \emph{ex vivo} samples. Performance is lower than on \emph{in vivo} data but our models still outperform \texttt{RAC} on almost all metrics.

\begin{table*}
\caption{Evaluation on \ToneWoneMM{} scans (except where noted) from the test and generalization sets. Metrics \gls{assd}, \gls{hd90}, and $\sigma(E)$ (standard deviation of edge lengths) are in mm; TQ (triangle quality; from \cref{eq:tri_quality}) is unitless; SIF and IF are in \%. Please note that surfaces produced by \gls{rac} are postprocessed with FreeSurfer to, for example, remove \gls{sif}. The best value within each dataset is highlighted. $n$ denotes the combined size of the datasets in each section.}
\begin{tabular}{p{1em} @{} l S[table-format=1.3] S[table-format=1.3] S[table-format=1.3] S[table-format=1.3] S[table-format=1.3] S[table-format=1.3] S[table-format=1.3] S[table-format=1.3] S[table-format=1.3] S[table-format=1.3] S[table-format=1.3]}
    \toprule
    && \multicolumn{5}{c}{\textbf{White matter}} & \multicolumn{5}{c}{\textbf{Gray matter}} & \textbf{WM-GM}\\
    && {ASSD} & {HD90} & {SIF} & {$\sigma(E)$} & {TQ} & {ASSD} & {HD90} & {SIF} & {$\sigma(E)$} & {TQ}  & {IF} \\
    \cmidrule(r){3-7} \cmidrule(l){8-12} \cmidrule(l){13-13}
    \multicolumn{13}{l}{\textbf{Test} ($n=490$)} \\
    \multicolumn{2}{l}{Ours} &&&&&&&&&&& \\
    & T1w-1mm      & \textbf{0.143} & 0.346 & \textbf{0.001} & 0.143 & 0.925 & 0.197 & 0.480 & 0.004 & \textbf{0.202} & 0.889 & \textbf{0.005} \\
    & synth-1mm    & 0.232 & 0.501 & \textbf{0.001} & 0.145 & 0.929 & 0.305 & 0.685 & \textbf{0.001} & 0.212 & \textbf{0.894} & 0.026 \\
    & synth-random & 0.282 & 0.622 & \textbf{0.001} & \textbf{0.142} & \textbf{0.930} & 0.366 & 0.816 & \textbf{0.001} & 0.221 & 0.892 & 0.016 \\
    \multicolumn{2}{l}{DeepSurfer}  & 0.152 & \textbf{0.331} & 0.004 & 0.172 & 0.891 & \textbf{0.177} & \textbf{0.392} & 0.016 & 0.323 & 0.774 & 0.024 \\
    \multicolumn{2}{l}{V2C-Flow}    & 0.185 & 0.427 & 1.194 & 0.351 & 0.740 & 0.245 & 0.605 & 2.359 & 0.383 & 0.748 & 0.070 \\
    &&&&&&&&&&&& \\

    \multicolumn{13}{l}{\textbf{Generalization} ($n=1287$)} \\
    \multicolumn{2}{l}{Ours} & &&&&&&&&&& \\
    & T1w-1mm      & 0.150 & 0.356 & \textbf{0.001} & 0.224 & 0.924 & 0.186 & 0.417 & 0.006 & \textbf{0.290} & 0.881 & \textbf{0.004} \\
    & synth-1mm    & 0.246 & 0.515 & \textbf{0.001} & \textbf{0.218} & \textbf{0.929} & 0.300 & 0.694 & \textbf{0.001} & \textbf{0.290} & 0.888 & 0.026 \\
    & synth-random & 0.298 & 0.647 & \textbf{0.001} & \textbf{0.218} & \textbf{0.929} & 0.365 & 0.865 & \textbf{0.001} & 0.303 & \textbf{0.889} & 0.017 \\
    \multicolumn{2}{l}{DeepSurfer}  & \textbf{0.143} & \textbf{0.316} & 0.002 & 0.249 & 0.897 & \textbf{0.175} & \textbf{0.384} & 0.007 & 0.414 & 0.781 & 0.017 \\
    \multicolumn{2}{l}{V2C-Flow}    & 0.182 & 0.410 & 1.164 & 0.411 & 0.739 & 0.223 & 0.549 & 4.195 & 0.422 & 0.740 & 0.080 \\
    &&&&&&&&&&&& \\

    \multicolumn{13}{l}{\textbf{Generalization} (\FLAIRfiveMM{}) ($n=199$)} \\
    \multicolumn{2}{l}{Ours} & &&&&&&&&&& \\
    & synth-random                  & \textbf{0.697} & \textbf{1.752} & 0.004 & \textbf{0.142} & \textbf{0.915} & \textbf{0.679} & \textbf{1.886} & 0.005 & \textbf{0.201} & \textbf{0.878} & \textbf{0.058} \\
    \multicolumn{2}{l}{\acrshort{rac}} & 0.940 & 2.376 & \textbf{0.001} & 0.291 & 0.894 & 0.926 & 2.689 & \textbf{0.002} & 0.466 & 0.750 & 0.135 \\
    &&&&&&&&&&&& \\

    \multicolumn{13}{l}{\textbf{BigBrain} ($n=1$)} \\
    \multicolumn{2}{l}{Ours} & &&&&&&&&&& \\
    & synth-1mm    & \textbf{0.428} & \textbf{1.326} & \textbf{0.000} & \textbf{0.263} & \textbf{0.891} & \textbf{0.351} & \textbf{0.839} & \textbf{0.000} & 0.314 & 0.868 & 0.220 \\
    & synth-random & 0.529 & 1.354 & 0.001 & \textbf{0.263} & 0.886 & 0.477 & 1.167 & 0.002 & \textbf{0.301} & \textbf{0.872} & \textbf{0.102} \\
    \multicolumn{2}{l}{\acrshort{rac}} & 0.813 & 2.346 & 0.001 & 0.286 & \textbf{0.891} & 0.557 & 1.167 & 0.002 & 0.437 & 0.782 & 0.124 \\
    &&&&&&&&&&&& \\

    \multicolumn{13}{l}{\textbf{Edlow} ($n=1$)}\\
    \multicolumn{2}{l}{Ours} & &&&&&&&&&& \\
    & synth-1mm    & \textbf{0.256} & \textbf{0.512} & \textbf{0.000} & \textbf{0.217} & \textbf{0.926} & 0.399 & 0.977 & 0.005 & 0.325 & 0.864 & 0.310 \\
    & synth-random & 0.308 & 0.824 & \textbf{0.000} & 0.221 & 0.922 & \textbf{0.383} & \textbf{0.859} & \textbf{0.000} & \textbf{0.315} & \textbf{0.871} & \textbf{0.072} \\
    \multicolumn{2}{l}{\acrshort{rac}} & 0.307 & 0.629 & 0.001 & 0.282 & 0.912 & 0.536 & 1.345 & 0.002 & 0.459 & 0.749 & 1.409 \\
    \bottomrule
\end{tabular}
\label{tab:eval_accuracy}
\end{table*}

\subsection{Surface Uncertainty}
\begin{figure}
\centering
\includegraphics[width=\linewidth]{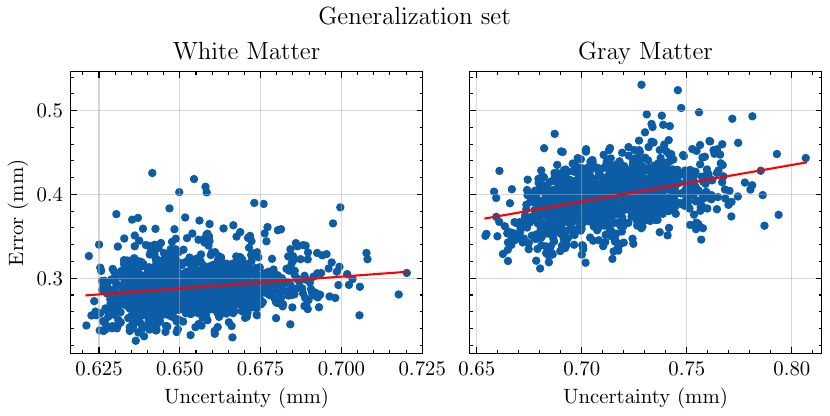}
\caption{Relationship between average uncertainty and surface reconstruction error for the generalization set using predictions from the T1w-1mm model. Pearson correlation cofficients are $r=0.16$ (\gls{wm}) and $r=0.46$ (\gls{gm}).}
\label{fig:uc_wholebrain_generalization_t1w_1mm}
\end{figure}

\begin{figure}
\centering
\includegraphics[width=\linewidth]{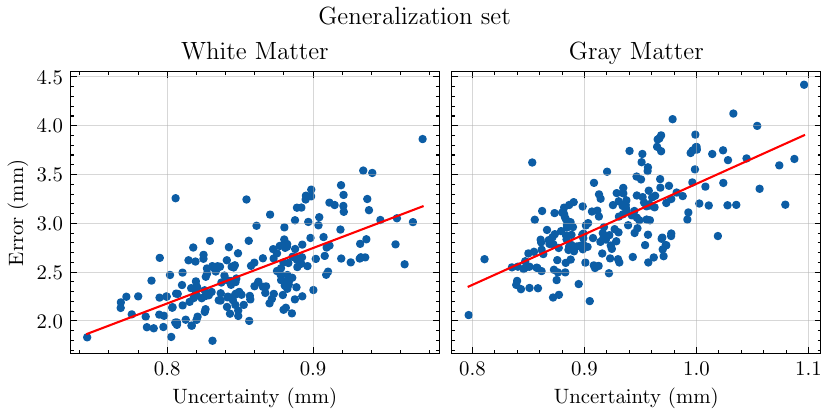}
\caption{Relationship between average uncertainty and surface reconstruction error for the low resolution generalization set using predictions from the synth-random model. Pearson correlation cofficients are $r=0.67$ (\gls{wm}) and $r=0.69$ (\gls{gm}).}
\label{fig:uc_wholebrain_generalization_flair_5mm}
\end{figure}

\Cref{fig:uc_wholebrain_generalization_t1w_1mm} shows the relationship between uncertainty and surface reconstruction error averaged over the entire cortex for each subject on \ToneWoneMM{} scans from the generalization set, respectively. For the \gls{wm} surface, we only see weak correlation ($r=0.16$), whereas the effect is stronger for the \gls{gm} surface ($r=0.46$); likely because the uncertainties and errors are generally small for \gls{wm} compared to \gls{gm} (in the supplementary material we show this relationship for the test set where we get correlations of $r=0.22$ and $r=0.74$).
However, in \cref{fig:uc_wholebrain_generalization_flair_5mm}, which shows the same for the low resolution \FLAIRfiveMM{} scans from the \gls{adni} GO-2 dataset, we see a moderate to strong correlation between uncertainty and error ($r=0.67$ for \gls{wm} and $r=0.69$ for \gls{gm}).

\begin{figure}
\centering
\includegraphics[width=\linewidth]{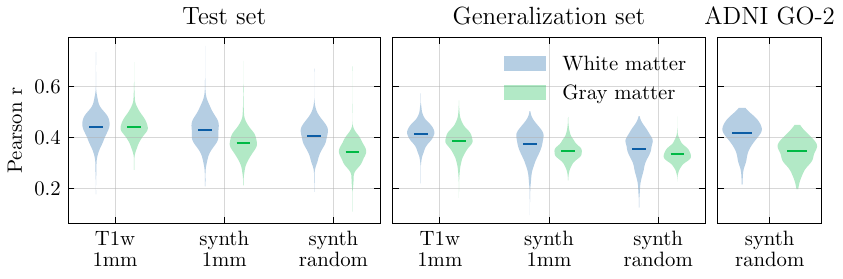}
\caption{Correlation between uncertainty estimate and smoothed surface errors within subjects. The bar in the violin plot indicates the mean of the distribution.}
\label{fig:uc_error_correlation}
\end{figure}

\Cref{fig:uc_error_correlation} shows distributions of vertex-wise correlations (over cortex per subject) between uncertainty and error. The correlations are modest (average values around $0.35-0.45$) and generally better for \gls{wm} and for the model trained specifically on \ToneWoneMM{} data.

\begin{figure}
    \centering
    \includegraphics[width=\linewidth]{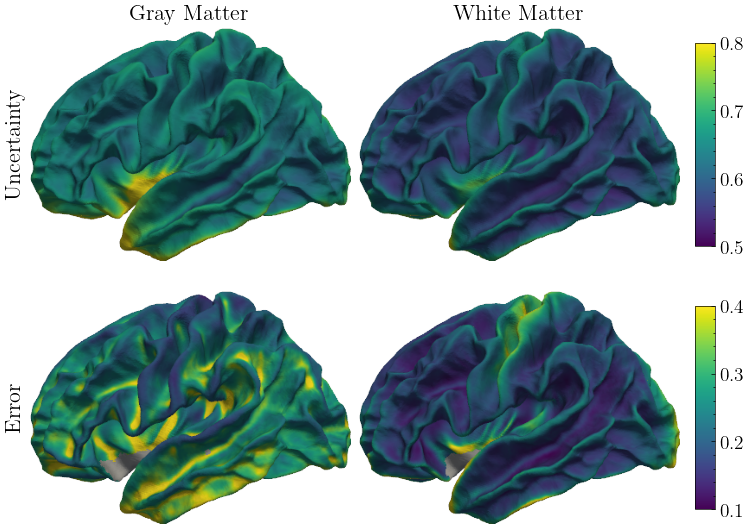}
    \caption{Average uncertainty and error over subjects from the generalization set mapped to \emph{fsaverage}. Surface predictions performed using the T1w-1mm model on \ToneWoneMM{} scans. The units are \si{\milli\meter}.}
    \label{fig:uc_fsaverage_generalization}
\end{figure}

\begin{figure}
    \centering
    \includegraphics[width=\linewidth]{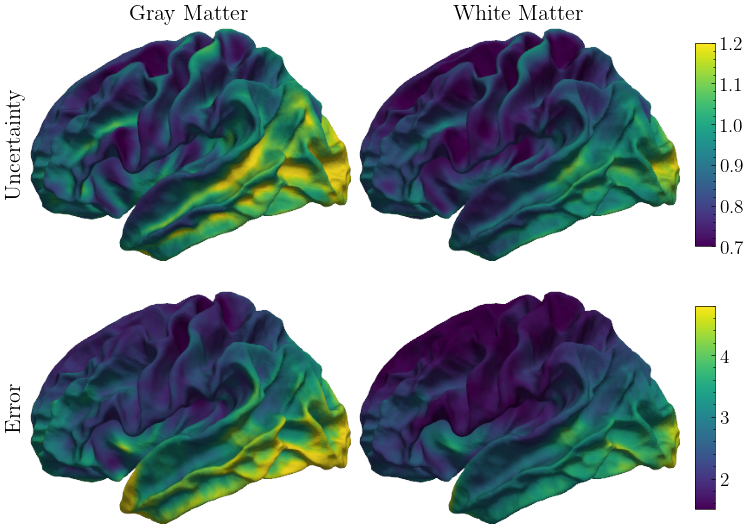}
    \caption{Average uncertainty and error over subjects from the \gls{adni}-GO2 test set mapped to \emph{fsaverage}. Surface predictions performed using the synth-random model on \FLAIRfiveMM{}. The units are \si{\milli\meter}.}
    \label{fig:uc_fsaverage_generalization_low}
\end{figure}

\Cref{fig:uc_fsaverage_generalization} shows mean uncertainty estimates and reconstruction errors for the T1w-1mm model (on the generalization set) mapped to \emph{fsaverage}. For the \gls{wm} surface, we see very similar patterns for both quantities as they are large on gyral crowns, in deep temporal areas, and in superior areas (around M1 and S1). For the \gls{gm} surface, the quantities also show similar patterns although the error is generally more pronounced around sulcal fundi than the uncertainty estimate. (In the supplementary material we show this data for the test set where we find patterns that are generally consistent with those from the generalization set.)
\Cref{fig:uc_fsaverage_generalization_low} shows the same but for the synth-random predictions on the \FLAIRfiveMM{} scans. We see increased error and uncertainty in occipital areas which are very difficult to reconstruct on low resolution scans due to the presence of many small sulci and gyri.

\subsection{Spherical Registration}
\begin{figure}
    \centering
    \includegraphics[width=\linewidth]{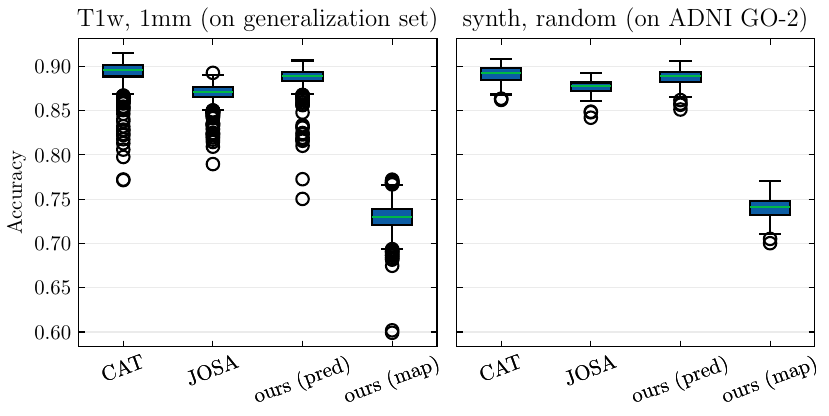}
    \caption{Boxplots of accuracy scores across all regions of the Desikan-Killiany atlas \cite{desikan2006}.
    (left) Spherical registration of surfaces predicted on \ToneWoneMM{} scans from the generalization set using the model \gls{t1w}-1mm.
    (right) Spherical registration of surfaces predicted on \ToneWoneMM{} scans from the \gls{adni} GO-2 dataset using the model synth-random.
    The reference against which the accuracy score is compute is the default spherical registration pipeline from FreeSurfer. The method denoted ``map'' uses coordinates obtained by spherical registration of our template to \emph{fsaverage}.}
    \label{fig:sphere-reg-acc}
\end{figure}

\Cref{fig:sphere-reg-acc} shows boxplots of classification accuracy computed by mapping the atlas from \cite{desikan2006} to each subject using different approaches. On the left, registrations are computed from surfaces estimated on \ToneWoneMM{} scans from the generalization set using our T1w-1mm model. On the right, registrations are computed from surfaces estimated on \FLAIRfiveMM{} scans from the \gls{adni} GO-2 dataset \emph{using our synth-random model}. In both figures, we see that \gls{cat12} performs best (mean accuracy of 0.894) followed by our prediction (0.887). JOSA has a slightly lower accuracy (0.867) but fewer outliers whereas the direct mapping approach generally performs poorly (0.730).
For all methods, we see a larger variability in classification accuracy when applied to surfaces predicted on \ToneWoneMM{} scans. Accuracy scores for different cortical regions are included in the supplementary material.

\section{Discussion}
\label{sec:discussion}
We have presented a \gls{dl} model that predicts (1) cortical surfaces, (2) an estimate of the uncertainty associated with location of each vertex on the surfaces, and (3) coordinates on the \emph{fsaverage} sphere corresponding to each vertex. We trained three different models: one on \ToneWoneMM{} scans, one on \qtyproduct[product-units = power]{1x1x1}{\milli\meter} synthetic scans, and one on synthetic scans of random resolution. The models trained on synthetic data can be applied to \gls{mri} scans of any contrast. Additionally, the latter model also handles scans with low resolution and anisotropic voxels.

\subsection{Surface Accuracy and Quality}
In terms of surface reconstruction accuracy, our T1w-1mm model performs similarly to (or better than) other \gls{dl} methods and the errors are similar across test and generalization set suggesting that it generalizes well to unseen data.
The synth-random model performs better than \gls{rac} and we note that it is significantly faster taking only about \qty{1}{\second} on a GPU whereas \gls{rac} has a runtime of a few hours (admittedly, \gls{rac} returns everything that \texttt{recon-all} does, however, the runtime is dominated by surface postprocessing).

\begin{figure}
\centering
\includegraphics[width=\linewidth]{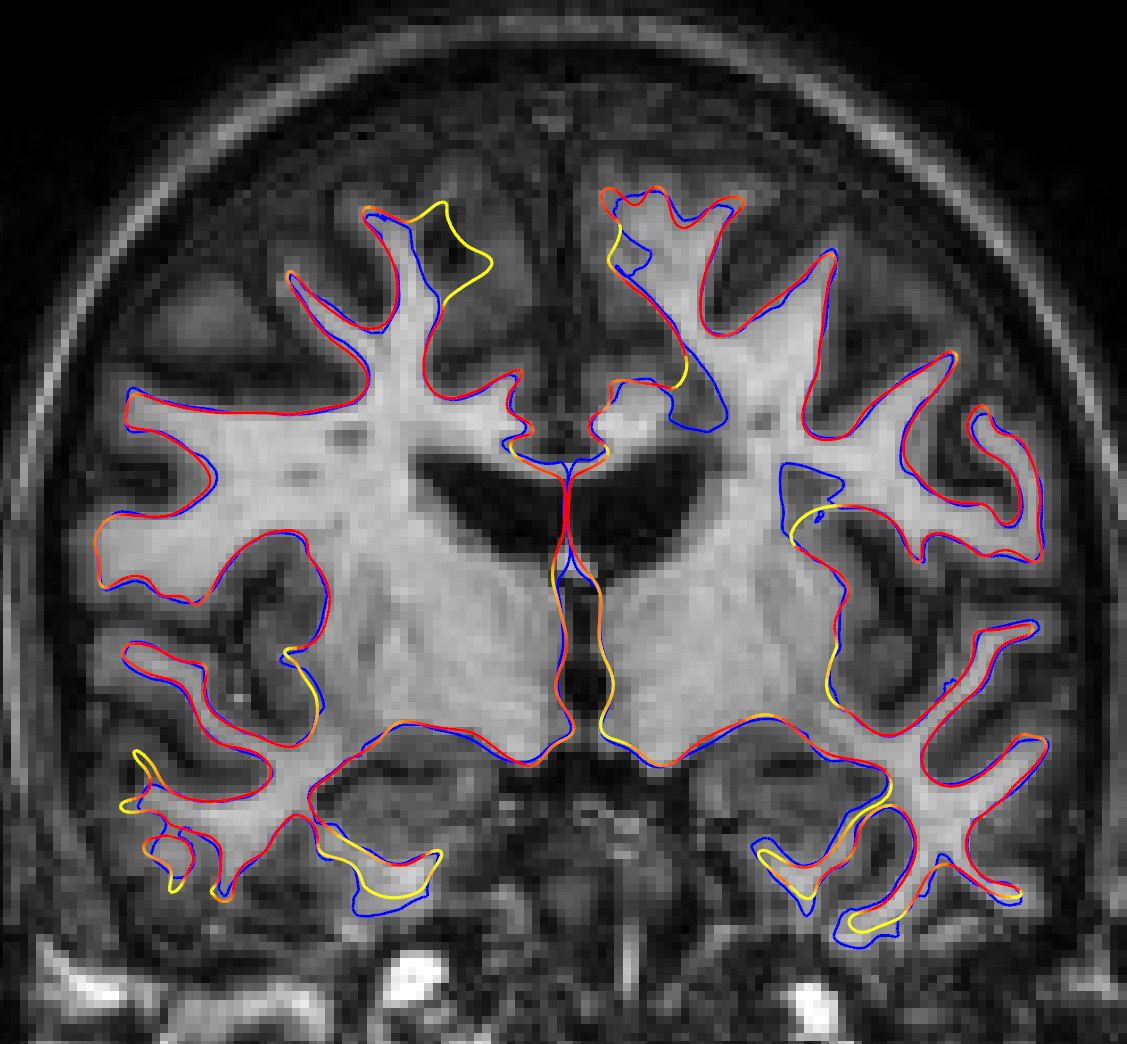}
\caption{\Gls{wm} surface reconstructions (FreeSurfer and our synth-random model) on the subject from \gls{isbi2015} with the highest reconstruction error. This illustrates that FreeSurfer (blue) is sensitive to \gls{wm} lesions (hypointensities) whereas the synth-random model (red-yellow where red and yellow corresponds to low and high uncertainty as estimated by the model, respectively) is less so. Additionally, we see that the estimated uncertainty highlights these difficult or problematic areas (even if the surface prediction might be correct).}
\label{fig:bad_fs_and_uc_lesion}
\end{figure}

We note that the lack of ground truth surfaces against which to quantify the error warrants a certain level of caution when interpreting these metrics. In particular, the common strategy of using FreeSurfer as reference, which can only be considered a silver standard in this regard, may sometimes be problematic. This becomes increasingly relevant as \gls{dl} models get more accurate and sometimes outperform FreeSurfer. When visually inspecting the results, we have seen many examples where the estimate of a particular \gls{dl} model is (locally) better than FreeSurfer (e.g., around a \gls{wm} hypointensities). We illustrate this in \cref{fig:bad_fs_and_uc_lesion} and it is our impression that this is a general feature of our models compared to FreeSurfer, namely that they tend to be more robust to abnormalities in the scans, perhaps because they have learned priors over shapes of sulci and gyri. On the other hand, it is also possible to find areas where FreeSurfer performs best and quantifying these findings is not trivial.
Thus, whether an \gls{assd} of \qty{0.13}{\milli\meter} is actually better than an \gls{assd} of \qty{0.15}{\milli\meter} may not be easy to say. This ambiguity is not a major concern when quantifying the performance on low resolution scans since a surface estimated from a \FLAIRfiveMM{} scan will almost surely be less accurate than a FreeSurfer prediction on an accompanying \ToneWoneMM{} scan.


From \cref{tab:eval_accuracy}, it is evident that the models trained on synthetic data generally perform worse when applied to \gls{t1w} scans than a model trained on this particular contrast. Consequently, the comparison to FreeSurfer-based ground-truth is less of an issue for the synthetically trained models. To explore why we see this performance gap, we have tried to explicitly model partial volume effects during data synthesis (using signed distances between a voxel and cortical surfaces; see \cite{nielsen2026} for a more detailed description). However, this did not make a difference compared to the approach used here, where we simply rely on image smoothing to create partial volume effects. Another reason for this performance gap could be differences in general image statistics (e.g., unrealistic noise model), but this is beyond the scope of this work. However, it should be noted that the performance is still well below the resolution of the scans.

In addition to high anatomical accuracy, we see from \cref{tab:eval_accuracy} that our methods produce high quality meshes; both in terms of triangle shape and size as well as low number of intersecting triangles. We visualize the surface meshes for the different models for a randomly selected sulcus in the supplementary material. This may be an attractive feature in many scenarios. For example, interpolating layers between \gls{wm} and \gls{gm} surfaces will be straightforward if the boundary surfaces are smooth. Likewise, high quality meshes may lend themselves more easily to subsequent mesh processing (e.g., remeshing) and improve numerical stability if used in for example finite element modeling. 

In \cref{tab:eval_accuracy}, we also show the results from applying our models to two \emph{ex vivo} cases, namely data from the BigBrain project and the MRI scan from Edlow et al. \cite{edlow2019}. The reference surfaces were obtained from \cite{paquola2026intracortical} and using the method from \cite{zeng2024}, respectively. Performance is decreased compared to the \emph{in vivo} data, however, it is still acceptable and demonstrates the flexibility of the models, and the approach in general, despite \emph{ex vivo} scans not being the focus of the current work. We speculate that by tweaking the augmentation scheme slightly towards \emph{ex vivo} cases, the models should be able to handle these cases much better. For example, our current models were trained exclusively on full-brain scans, but single hemisphere augmentation is straightforward.


\subsection{Surface Uncertainty}
We evaluated the value of the uncertainty estimates produced by our models by correlating them with observed surface placement errors. We found that (1)~uncertainty averaged over the entire cortex is a good predictor of overall surface accuracy when applied to low resolution scans, whereas (2)~vertex-wise uncertainty estimates are able to highlight areas of potentially problematic surface reconstruction within a subject (especially for the \gls{gm} surface) on high resolution scans.

Evaluating the predicted uncertainties of the models is difficult since we do not have a target for this quantity which is why we do it indirectly by correlating uncertainty and observed reconstruction error. However, as discussed above, FreeSurfer surfaces can only be considered a silver standard in this regard making it difficult to determine if
\begin{enumerate}
    \item low uncertainty but high error is actually a false negative (where our model fails to recognize a potential error) or if our model is actually correct whereas FreeSurfer is not (thus justifying a low uncertainty), and
    \item high uncertainty but low error is actually a false positive because both models are correct (although if both models happen to be correct despite high level of noise in the image this might still be okay) or both models are wrong in a similar way, hence justifying high uncertainty.
\end{enumerate}
This ambivalence is particularly problematic when looking at within-subject correlations because the vast majority of uncertainty and error values are low meaning that even a few occurrences of the cases will be detrimental to the overall correlation if their interpretation as false negative/positive is not warranted (otherwise, low correlation is the correct result).

\begin{figure*}
    \centering
    \begin{subfigure}[t]{0.32\linewidth}
        \centering
        \includegraphics[width=\linewidth]{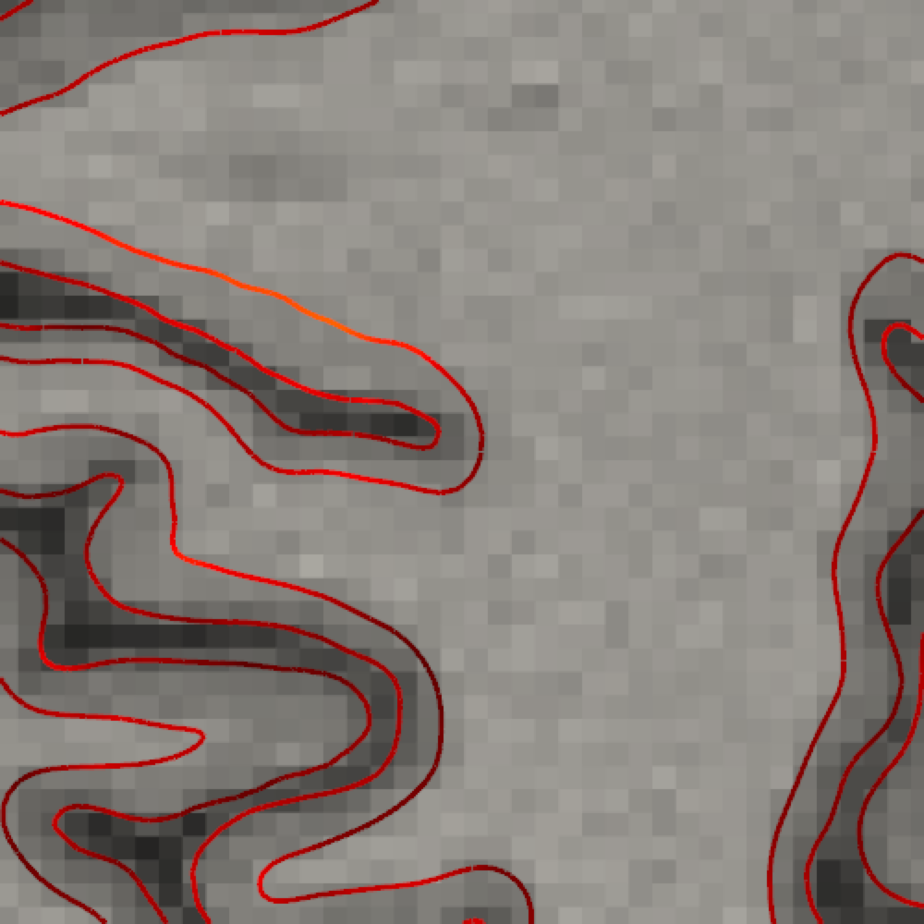}
        \caption{No (added) noise.}
        \label{fig:uc_t1w}
    \end{subfigure}
    \begin{subfigure}[t]{0.32\linewidth}
        \centering
        \includegraphics[width=\linewidth]{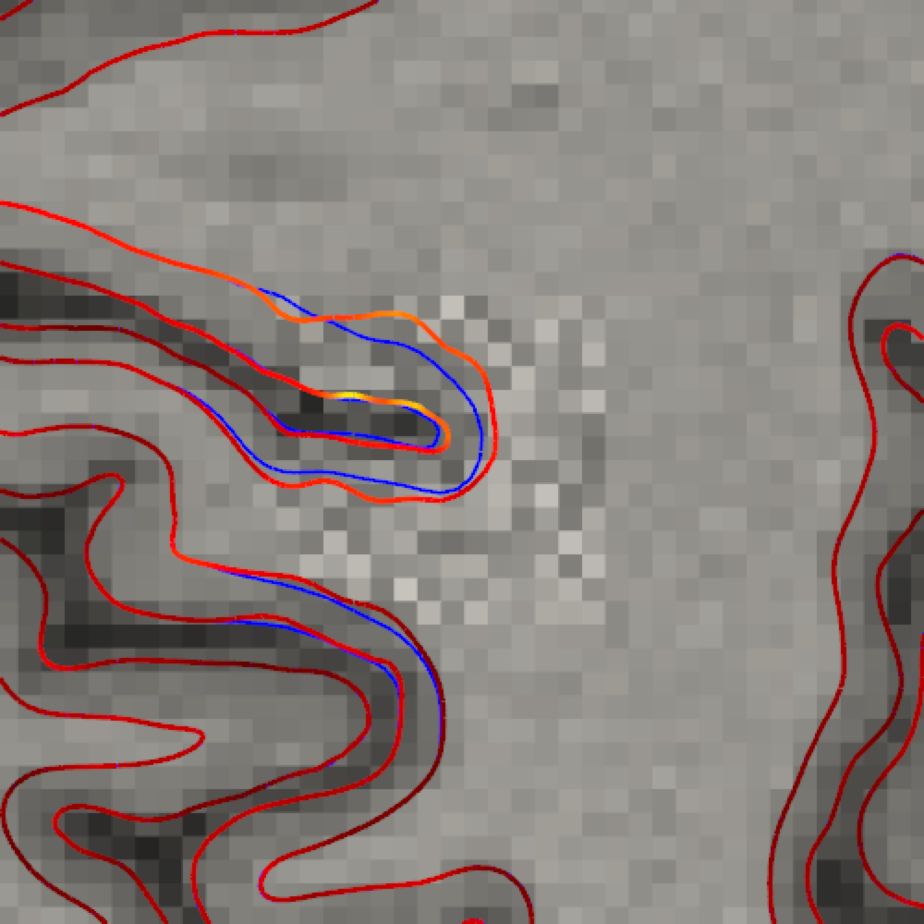}
        \caption{Low noise.}
        \label{fig:uc_t1w_noise_low}
    \end{subfigure}
    \begin{subfigure}[t]{0.32\linewidth}
        \centering
        \includegraphics[width=\linewidth]{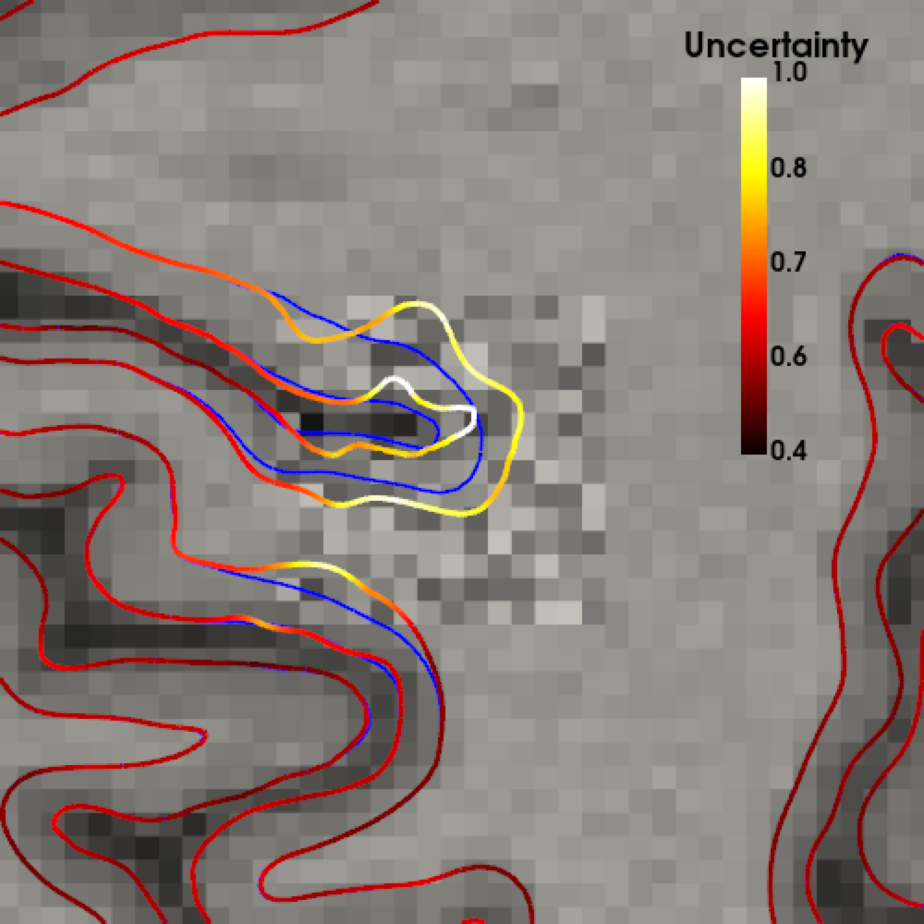}
        \caption{High noise.}
        \label{fig:uc_t1w_noise_high}
    \end{subfigure}
    \caption{Surface reconstructions on \ToneWoneMM{} scans with (\subref{fig:uc_t1w}) no added noise, (\subref{fig:uc_t1w_noise_low}) low amount of local noise, or (\subref{fig:uc_t1w_noise_high}) high amount of local noise. Black (yellow/white) indicates low (high) values of uncertainty. Blue is the surface reconstruction from (\subref{fig:uc_t1w}) with no overlay. Noise is sampled from a uniform distribution and added in a window of size \qtyproduct[product-units = power]{14x14x14}{voxels}.}
    \label{fig:uncertainty_response_to_local_noise}
\end{figure*}

An additional feature of the uncertainty estimates of our models is that they are generally biased towards high uncertainty in certain regions (e.g., insula, temporal pole area, orbitofrontal area, occipital area) as is clear from \cref{fig:uc_fsaverage_generalization}. The reason for this probably depends on the region. For example, the contrast in \gls{t1w} scans is often not good in insula, orbitofrontal and temporal areas and so \texttt{recon-all} may be less reliable here (thus increasing arbitrary variation in the training data), whereas the occipital area is simply difficult because it consists of many small gyri and sulci.
On the other hand, from \cref{fig:bad_fs_and_uc_lesion} it is also clear that the uncertainty estimates are sensitive to local features in the image. To illustrate this, \cref{fig:uncertainty_response_to_local_noise} demonstrates this in a controlled setting where we show that increasing the amount of simulated noise in a small area of an image increases the (local) uncertainty estimate of the model.
Indeed, from looking at uncertainty overlaid on the surfaces for many subjects, it is our general impression that the uncertainty estimate is almost always high in areas that have poor contrast or are somehow abnormal (e.g., pathology). In addition, when severe reconstruction errors occur, we have seen that this is often highlighted by the uncertainty estimate.

Lastly, we acknowledge the importance for models to provide well-calibrated uncertainty estimates. This is an active area of research \cite{zhang2026} and while we have not performed a formal calibration of our uncertainty estimates, mainly because we have no ground truth data against which to perform such calibration, we believe that our results demonstrate the usefulness of our uncertainty estimates, e.g., for quality control or downstream applications.



\subsection{Surface Registration}
Our methods generally provide accurate spherical registrations that are similar to those from \gls{cat12} and JOSA. To evaluate the quality of our predicted spherical registration, we used the standard FreeSurfer pipeline as reference for calculating vertex-wise accuracy. Although manual inspection of a subset of the registrations suggested that it performs well in the vast majority of cases, and thus is suitable as reference, it is worth noting that this is not a gold standard.

From \cref{fig:sphere-reg-acc} (left), we see a few more cases with low accuracy for \gls{cat12} and our method compared to JOSA. These outliers mostly reflect disagreement on where to locate the pre- and postcentral gyri. Interestingly, \gls{cat12} and our approach generally agree, although manual inspection suggested that FreeSurfer is correct in these cases. Given that JOSA uses the same rigid transformation for the initial registration as the standard pipeline from FreeSurfer, it is not surprising that we do not find major disagreements between these methods. On the other hand, the average error of JOSA is higher than \gls{cat12} and our predictions.

Additionally, we observed greater variability when the registration methods were applied to \ToneWoneMM{} scans compared to \FLAIRfiveMM{} scans. This is consistent across all methods and is likely because the surfaces predicted from the former scans generally capture more details than the latter, hence increasing the possibility of the methods settling in different local minima.

To our knowledge, the only other \gls{dl} method that does simultaneous cortical surface reconstruction and spherical registration was presented by \cite{rickmann2025}. However, since we were unable to find parameters for this model, we have not been able to do a proper comparison. Additionally, since the registration is tied to the predicted surface, it is not possible to apply either method to the surface prediction of the other.
In \cite{rickmann2025}, the authors find average Dice scores of approximately 0.92 which is impressive, although not directly comparable to our metric of parcellation accuracy. However, since resampling a surface onto the \emph{fsaverage} template tends to degrade smoothness and triangle quality (e.g., shear triangles; see supplementary material for an example), we speculate that the mesh quality may be negatively affected when driving the training exclusively with a vertex correspondence loss.

Finally, our approach and that of \cite{rickmann2025} are different from most other \gls{dl} methods that are trained only to perform spherical registration. This makes dedicated registration methods (such as JOSA) versatile, but also dependent on other tools, e.g., for spherical inflation and feature (curvature, sulcal depth, etc.) estimation. The spherical mapping step, in particular, can be time-consuming with FreeSurfer.



\subsection{Limitations}
There are a number of limitations relevant to this work that are worth highlighting.
Regarding the models themselves, we note the following:
\begin{itemize}
    \item The prediction accuracy and surface quality from our models are on average high, but there are no guarantees or upper bounds on the errors, and the models may fail if applied to data that are outside of the training data distribution. For example, although we have found that our models are quite robust to white matter hypo- and hyperintensities, the accuracy of the surfaces in the presence of (major) cortical abnormalities or artifacts remains to be evaluated.
    \item Our models expect an input image with \qtyproduct[product-units = power]{1x1x1}{\milli\meter} voxels as inputs. If the input scan has a lower resolution, it will be upsampled, however, if it has a \emph{higher} resolution it will need to be downsampled. Consequently, the networks will not benefit from sub-millimeter resolution.
    \item In general, the data synthesis pipeline used for training the ``synth'' models does not generate realistic \gls{mri} scans. Rather, we rely on generating a sufficiently broad distribution of data that will encompass the cases encountered in reality. This approach has been successfully used in many other applications (e.g., \cite{billot2023,gopinath2025,gopinath2024a}). However, we observed that in some cases, surfaces may be consistently over- or underestimated by a fraction of a voxel. We have not been able to pin down the exact reason(s), but it is evident that the noise model used here (Gaussian noise) does not reflect the true noise characteristics in \gls{mri} scans.
\end{itemize}
Another set of limitations relate not to the models themselves but to their evaluation because of the lack of a ground truth data:
\begin{itemize}
    \item As mentioned above, surface estimates from FreeSurfer cannot be considered ground truth and therefore differences in \gls{assd} below a certain threshold may be meaningless. At the very least, one should be cautious about putting too much emphasis on small differences between models. We excluded subjects where FreeSurfer predictions were very bad. However, caution has to be exercised with this approach as we may be excluding scans in a biased way (e.g., cases that are somehow at the tail of the distribution) thus limiting the generalization performance of the trained model. However, this may not be a major concern as we found the generalization performance of our models to be on par with the performance on the test set.
    \item We do believe that the predicted uncertainty estimates can be useful for example for quality control purposes. However, as we have tried to stress, it can be difficult to quantify its exact accuracy and value (this ties back to the previous point).
    \item Regarding the registration to \emph{fsaverage}, FreeSurfer is also not a ground truth although in our experience it tends to perform well.
\end{itemize}
Finally, to ensure stability during training on synthetic data, we chose to pretrain the models to match the features from the \gls{t1w} model. One limitation of this approach is that the final number of feature channels from the UNet must match. An alternative strategy is to stabilize the training using a matched vertex loss in the initial training phase \cite{hoopes2022} although this is generally slower.

\section{Conclusions}
Here we have presented three \gls{dl} models, which, given an input brain scan, return (1) cortical surfaces, (2) uncertainty estimates, and (3) spherical registration to \emph{fsaverage} associated with said cortical surfaces. We have trained a model specifically on \ToneWoneMM{} scans but also a model that can handle scans of any contrast and resolution (e.g., clinical scans). Our approach is very fast and, if available on the control PC of a scanner, could allow immediate surface reconstruction with uncertainty estimates as well as mapping of cortical atlases on a subject during a scanning session to help identify areas of potential interest. 

We provide a command line interface which takes an input scan and produces the above output without the need for manual preprocessing or registration.
The code is available at \url{https://github.com/simnibs/brainnet}.


\printcredits

\section*{Acknowledgements}
JDN and OP were supported by the Lundbeck Foundation (R360–2021–39).
AD was supported by the National Institutes of Health (NIH) (R01EB033773).
AT was supported by the Lundbeck Foundation (R313-2019-622), the German Research Foundation (DFG grants TH 1330/6-1 and TH 1330/7-1, part of Research Unit FOR 5429 ``MeMoSLAP''), NIH (1RF1MH117428-01A1) and European Union's Horizon Europe research and innovation programme (grant agreement No. 101071008).
JEI was supported by NIH -- BRAIN Initiative (1RF1MH123195, 1UM1MH130981), NIH -- National Institute of Aging (1R01AG070988,\\ 1RF1AG080371, 1R21NS138995), and NIH -- NIBIB \\(1R01EB031114).
\bibliographystyle{plain}
\bibliography{references}

\end{document}


\maketitle
\section{Methods}

\subsection{Model Training}
\subsubsection{Domain Randomization}
The following is an extended description of the same section in the main text.

We generate synthetic \gls{mr} images by sampling intensity values from a normal distribution conditioned on a label image, $L$, where $L(i)=k$ denotes the label of the $i$th voxel. Specifically, we sample a mean, $\mu$, for each label, $k$, and draw an image, $I$, according to
\begin{align}
    \mu_k   &\sim U\left( 25, 200 \right) \\
    I(i)    &= \mu_{L(i)}
\end{align}
where $U$ is the uniform distribution. We apply smoothing with a normal kernel whose standard deviation is drawn from $U(0.5, 1.2)$ to simulate partial volume effects and avoid sharp borders between regions with different labels. Finally, we sample a standard deviation, $\sigma$, for each label and add normally distributed noise according to
\begin{align}
    \sigma_\text{min} &\sim U(0, 6) \\
    \sigma_k    &\sim U( \sigma_\text{min}, \sigma_\text{min} + 5 ) \\
    I(i)        &\sim \max\left( 0, N\left(I(i), \sigma_{L(i)}^2\right) \right).
\end{align}
We force a minimum contrast of 10 between \gls{wm}, \gls{gm}, and \gls{csf} to stabilize the training. 
Next, we apply a Gamma transformation whose coefficient is sampled from
\begin{align}
    g &\sim \Gamma(0.5, 0.5) \\
    \gamma &\sim \max(0.5,\min(g, 2))
\end{align}
and add a slowly varying bias field as done in~\cite{billot2023}
\begin{align}
    s &\sim U(0.01, 0.04) \\
    B_{lr} &\sim N_{sW \times sH \times sD}(0, 0.6) \\
    B &= \text{upsample} (B_{lr}) \\
    I &= I \times \exp{ B }.
\end{align}
Here, $W \times H \times D$ is the size of $I$ and $s$ is a scaling factor.
Finally, we apply resolution augmentation to simulate different scanning protocols~\cite{iglesias2023}. Specifically, we randomly select one of the following scenarios: (1) no augmentation (\qty{1}{\milli\meter\cubed} voxels), (2) clinical scans (thick slices of \qtyrange{2.5}{8.5}{\milli\meter} with \qty{1}{\milli\meter\squared} in-plane resolution), (3) axial low field sequence (thick axial slices of \qtyrange{5}{5.4}{\milli\meter}), (4) near-isotropic low field sequence (\qtyrange{2}{5}{\milli\meter} along each axis). To provide a consistent input to the network, images are always resampled to \qty{1}{\milli\meter\cubed} voxels using trilinear interpolation, and intensities are normalized to values between 0 and 1. 

To increase robustness to inaccurate template registration, we perturb the template vertex coordinates by randomly drawn affine paramters
\begin{align}
    \text{translation} &\sim U\left( -3, 3 \right) \\
    \text{rotation}    &\sim U\left( -5, 5 \right) \\
    \text{shear}       &\sim U\left( -0.05, 0.05 \right)\\
    \text{scale}       &\sim U\left( 0.95, 1.05 \right)    
\end{align}
with a probability of 0.5.

\subsection{Implementation}
\Cref{alg:wm_est,alg:gm_est} present pseudo-code for the white matter and gray matter (pial) surface are generation.

\begin{algorithm}
\caption{White matter surface estimation. As the functions $g_n$, we use a UNet-like architecture of graph convolutional layers (see main text for details). \textsc{sample}$(A,B)$ performs trilinear interpolation of the volume $A$ at coordinates $B$, \textsc{proj} projects the vertices on the sphere, and \textsc{upsample} inserts a new vertex at the center of each edge whose value (position, uncertainty, etc.) is computed as the average of the vertices making up the edge.}
\label{alg:wm_est}
\begin{algorithmic}
    \Require $V(0)$ \Comment{Template positions}
    \Require $S(0)$ \Comment{Mapped spherical coordinates}
    \Require $F$ \Comment{Image features}
   
    \State $N \gets 7$ \Comment{Number of scales}
    \State $K \gets 2$ \Comment{Number of integration steps (per scale)}
    \State $h \gets 1/K$
    \State $V \gets V(0)$ \Comment{Vertex coordinates}
    \State $\sigma \gets \mathbf{0}$ \Comment{Uncertainty of vertex coordinates}
    \State $S \gets S(0)$ \Comment{Spherical coordinates of vertices}
    \For{$n = 1, \dots, N$}
        \For{$k = 1, \dots, K$} \Comment{Solve ODE}
            \State $F_V \gets \,$ \Call{sample}{$F$, $V$}
            \State $\begin{bmatrix} dV \; d\sigma \; dS \end{bmatrix}^\top \gets g_{n} (F_V) \,$ 
            \State $V \gets V + h \times dV$
            \State $\sigma \gets \sigma + h \times d\sigma$
            \State $S \gets S + h \times dS$    
            \State $S \gets \,$\Call{proj}{$S$}    
        \EndFor
        \If{$n < N$}
            \State $V \gets \,$ \Call{upsample}{$V$}
            \State $\sigma \gets \,$ \Call{upsample}{$\sigma$}
            \State $S \gets \,$ \Call{upsample}{$S$}
            \State $S \gets \,$ \Call{proj}{$S$}
        \EndIf
    \EndFor
    \State \Return $V$, $\exp{\sigma}$, $S$
\end{algorithmic}
\end{algorithm}

\begin{algorithm}
\caption{Gray matter surface estimation. As the function $g$, we use a sequence of linear layers and \gls{prelu} activation functions. \textsc{sample}$(A,B)$ performs trilinear interpolation of the volume $A$ at coordinates $B$.}
\label{alg:gm_est}
\begin{algorithmic}
    \Require $V_\text{WM}$ \Comment{White matter surface}
    \Require $\sigma_\text{WM}$ \Comment{White matter uncertainty}
    \Require $F$ \Comment{Image features}

    \State $K \gets 10$ \Comment{Number of integration steps}
    \State $h \gets 1/K$
    \State $V \gets V_\text{WM}$
    \State $\sigma \gets \log{\sigma_\text{WM}}$
    
    \For{$k = 1, \dots, K$} \Comment{Solve ODE}
        \State $F_V \gets \,$ \Call{sample}{$F$, $V$}
        \State $\begin{bmatrix} dV \; d\sigma \;\end{bmatrix}^\top \gets g (F_V) \,$ 
        \State $V \gets V + h \times dV$
        \State $\sigma \gets \sigma + h \times d\sigma$
    \EndFor
    \State \Return $V$, $\exp{\sigma}$
\end{algorithmic}
\end{algorithm}

\section{Results}
\subsection{Surface Accuracy and Quality}
\Cref{tab:eval_accuracy_assd} shows the \gls{assd} for different datasets. As described in the main text, we do not see clear differences between datasets (except ISBI2015) suggesting that the model predicts are reasonably reliable.

\begin{table}
\caption{Evaluation on \ToneWoneMM{} scans from the test set (top) and generalization set (bottom) using \gls{assd} in mm. DS (DeepSurfer), V2C (V2C-Flow). Contrast and resolution refers to the scans on which the models were trained.}
\begin{tabular}{l S[table-format=1.3] S[table-format=1.3] S[table-format=1.3] S[table-format=1.3] S[table-format=1.3] S[table-format=1.3] S[table-format=1.3] S[table-format=1.3] S[table-format=1.3] S[table-format=1.3]}
\toprule
& \multicolumn{5}{c}{\textbf{White matter}} & \multicolumn{5}{c}{\textbf{Gray matter}} \\
Contrast & \multicolumn{3}{c}{T1w} & \multicolumn{2}{c}{synth} & \multicolumn{3}{c}{T1w} & \multicolumn{2}{c}{synth} \\
Resolution & \multicolumn{3}{c}{1 mm} & {1 mm} & {rand} & \multicolumn{3}{c}{1 mm} & {1 mm} & {rand} \\
Model & {DS} & {Ours} & {V2C} & {Ours} & {Ours} & {DS} & {Ours} & {V2C} & {Ours} & {Ours} \\
\cmidrule(r){2-6} \cmidrule(r){7-11} 
ABIDE       & 0.153 & 0.150 & 0.211 & 0.247 & 0.291 & 0.210 & 0.239 & 0.328 & 0.353 & 0.409 \\
ADHD200     & 0.144 & 0.138 & 0.187 & 0.213 & 0.259 & 0.206 & 0.232 & 0.329 & 0.332 & 0.389 \\
ADNI3       & 0.197 & 0.179 & 0.198 & 0.273 & 0.344 & 0.171 & 0.186 & 0.200 & 0.293 & 0.367 \\
AIBL        & 0.155 & 0.153 & 0.192 & 0.248 & 0.300 & 0.165 & 0.178 & 0.192 & 0.295 & 0.356 \\
Buckner40   & 0.118 & 0.130 & 0.151 & 0.206 & 0.248 & 0.164 & 0.198 & 0.205 & 0.310 & 0.374 \\
Chinese-HCP & 0.127 & 0.130 & 0.162 & 0.203 & 0.260 & 0.155 & 0.190 & 0.232 & 0.277 & 0.335 \\
HCP         & 0.131 & 0.111 & 0.155 & 0.185 & 0.237 & 0.149 & 0.172 & 0.222 & 0.262 & 0.327 \\
OASIS3      & 0.165 & 0.152 & 0.184 & 0.253 & 0.303 & 0.167 & 0.175 & 0.193 & 0.298 & 0.358 \\
\midrule
COBRE       & 0.149 & 0.144 & 0.179 & 0.229 & 0.286 & 0.184 & 0.165 & 0.195 & 0.284 & 0.341 \\
ISBI2015    & 0.129 & 0.138 & 0.206 & 0.241 & 0.276 & 0.191 & 0.241 & 0.285 & 0.417 & 0.451 \\
MCIC        & 0.175 & 0.179 & 0.207 & 0.255 & 0.308 & 0.166 & 0.197 & 0.239 & 0.296 & 0.370 \\
AOMIC-ID1000 & 0.136 & 0.147 & 0.177 & 0.248 & 0.300  & 0.175 & 0.187 & 0.224 & 0.301 & 0.367  \\

\bottomrule
\end{tabular}
\label{tab:eval_accuracy_assd}
\end{table}

\subsubsection{Mesh Quality}
\begin{figure}
    \centering
    \includegraphics[width=\linewidth]{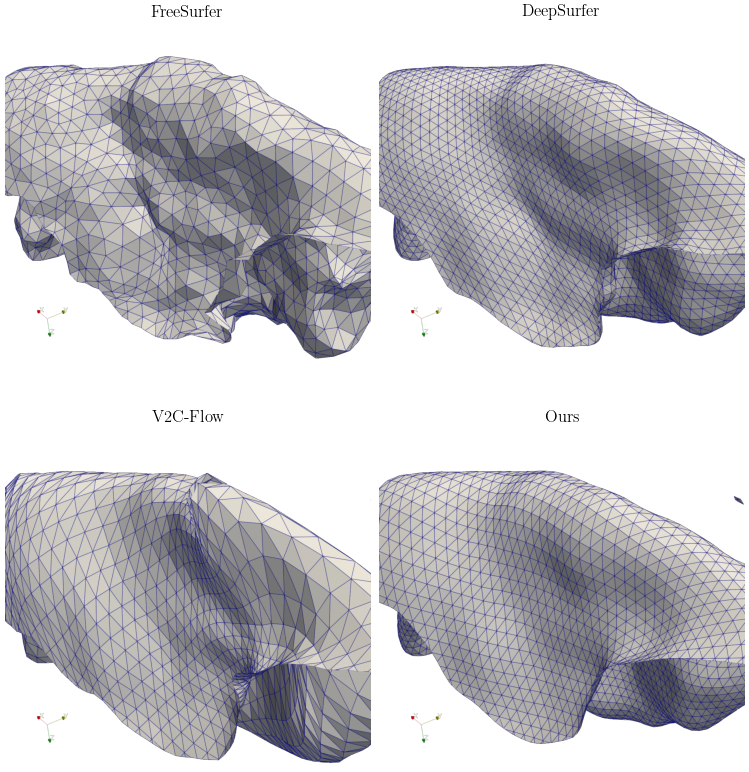}
    \caption{White matter surface. The view shows a sulcus from below.}
    \label{fig:mesh_q_white}
\end{figure}

\begin{figure}
    \centering
    \includegraphics[width=\linewidth]{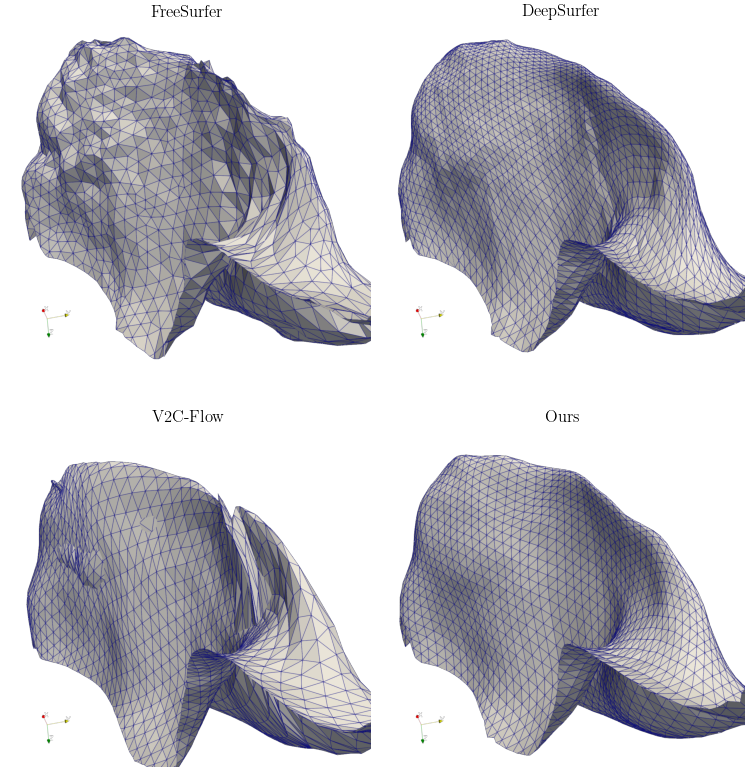}
    \caption{Gray matter surface. The view shows a sulcus from below.}
    \label{fig:mesh_q_gray}
\end{figure}

\Cref{fig:mesh_q_white,fig:mesh_q_gray} illustrate the triangulation quality of the different methods. We note that the surface meshes contain different number of vertices, namely 155,756 (FreeSurfer), 163,842 (V2C-Flow), and 245,762 (DeepSurfer, ours). Both figures show the same sulcus viewed from below. This is a difficult place to model the pial surface because of the high curvature. The particular subject and sulcus was randomly selected but we believe this is still a representative example (although the surface from V2C-Flow happens to contain self-intersections in this view).

\subsection{Uncertainty}
\begin{figure}
\centering
\includegraphics[width=\linewidth]{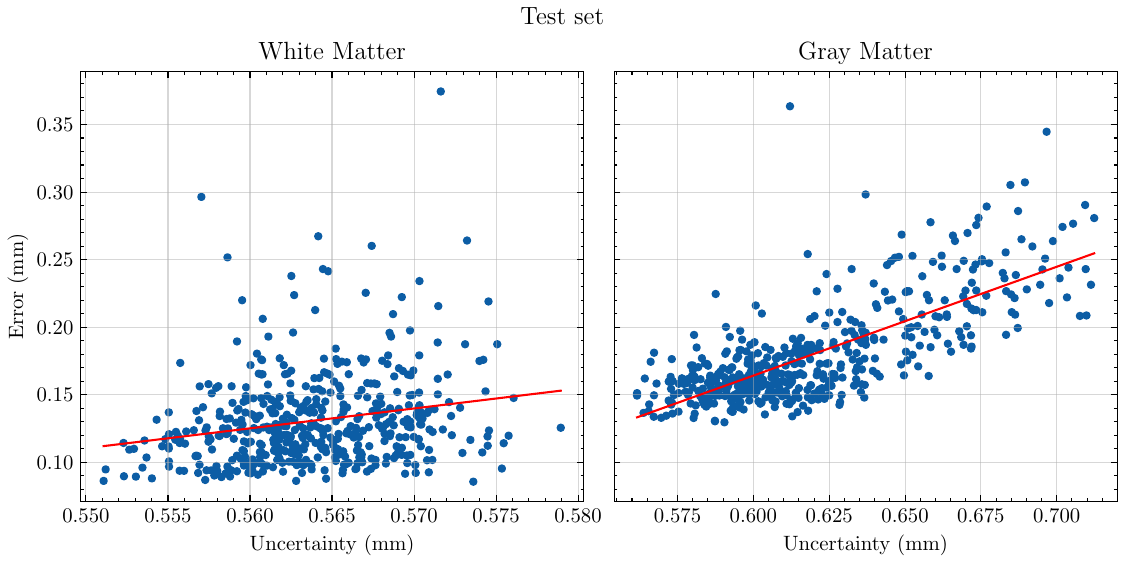}
\caption{Relationship between average uncertainty and surface reconstruction error for the test set using predictions from the T1w-1mm model. Pearson correlation coefficients are $r=0.22$ (\gls{wm}) and $r=0.74$ (\gls{gm}).}
\label{fig:uc_wholebrain_test}
\end{figure}

\Cref{fig:uc_wholebrain_test} shows the relationship between uncertainty and surface reconstruction error averaged over the entire cortex for each subject on \ToneWoneMM{} scans from the test set, respectively. For the \gls{wm} surface, we only see weak correlation ($r=0.22$), whereas the effect is stronger for the \gls{gm} surface ($r=0.74$).

\begin{figure}
    \centering
    \includegraphics[width=\linewidth]{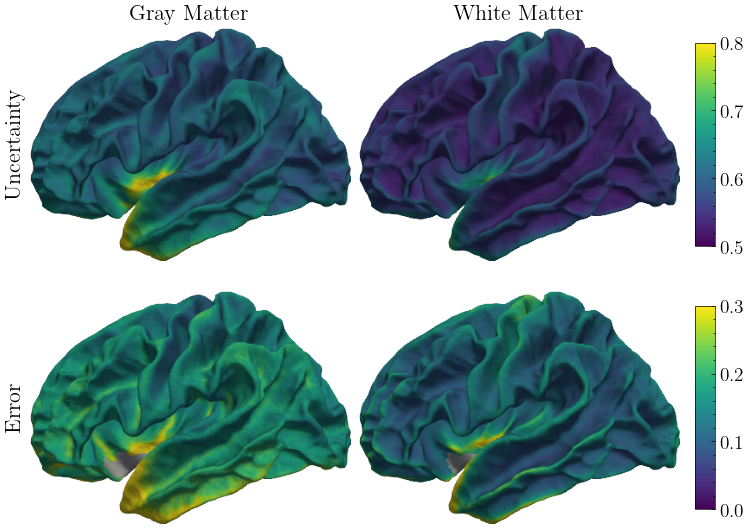}
    \caption{Average uncertainty and error over subjects from the test set mapped to \emph{fsaverage}. Surface predictions performed using the T1w-1mm model on \ToneWoneMM{} scans. The units are \si{\milli\meter}.}
    \label{fig:uc_fsaverage_test}
\end{figure}

\Cref{fig:uc_fsaverage_test} shows mean uncertainty estimates and reconstruction errors for the T1w-1mm model (on the test set) mapped to \emph{fsaverage}. For the \gls{wm} surface, we see very similar patterns for both quantities as they are large on gyral crowns, in deep temporal areas, and in superior areas (around M1 and S1). For the \gls{gm} surface, the quantities also show similar patterns although the error is generally more pronounced around sulcal fundi than the uncertainty estimate.

\subsection{Spherical Registration}
We show Dice scores per cortical region from spherical registrations estimated in two scenarios: \cref{tab:dice_region_t1w_1mm} shows results of surfaces estimated on \qtyproduct[product-units = power]{1x1x1}{\milli\meter} \gls{t1w} scans from the generalization set (see \cref{tab:dice_region_t1w_1mm}); \cref{tab:dice_region_flair_5mm} shows results of surfaces estimated on \qtyproduct[product-units = power]{1x1x5}{\milli\meter} \gls{flair} scans from the \gls{adni} GO-2 dataset.

\begin{table}
\caption{Classification accuracy per cortical region of the Desikan-Killiany atlas \cite{desikan2006}. The results are computed using on \qtyproduct[product-units = power]{1x1x1}{\milli\meter} \gls{t1w} scans from the generalization set (AOMIC-ID1000, COBRE, ISBI2015, MCIC). We use our model trained on \qtyproduct[product-units = power]{1x1x1}{\milli\meter} \gls{t1w} scans. The reference, against which the accuracy is computed, is the default spherical registration pipeline from FreeSurfer.}
\label{tab:dice_region_t1w_1mm}
\centering
\begin{tabular}{lSSSS}
\toprule
& {CAT} & {JOSA} & {ours (est)} & {ours (map)} \\
Region &  &  &  &  \\
\midrule
bankssts & 0.84 & 0.82 & 0.83 & 0.39 \\
caudalanteriorcingulate & 0.86 & 0.82 & 0.85 & 0.36 \\
caudalmiddlefrontal & 0.86 & 0.84 & 0.86 & 0.65 \\
cuneus & 0.87 & 0.75 & 0.82 & 0.70 \\
entorhinal & 0.73 & 0.65 & 0.73 & 0.32 \\
frontalpole & 0.39 & 0.25 & 0.38 & 0.02 \\
fusiform & 0.83 & 0.83 & 0.87 & 0.73 \\
inferiorparietal & 0.91 & 0.89 & 0.90 & 0.78 \\
inferiortemporal & 0.82 & 0.84 & 0.86 & 0.67 \\
insula & 0.92 & 0.93 & 0.92 & 0.72 \\
isthmuscingulate & 0.87 & 0.80 & 0.82 & 0.53 \\
lateraloccipital & 0.89 & 0.89 & 0.89 & 0.85 \\
lateralorbitofrontal & 0.92 & 0.87 & 0.92 & 0.70 \\
lingual & 0.89 & 0.89 & 0.88 & 0.82 \\
medialorbitofrontal & 0.89 & 0.83 & 0.88 & 0.64 \\
medialwall & 0.94 & 0.92 & 0.93 & 0.79 \\
middletemporal & 0.87 & 0.86 & 0.87 & 0.62 \\
paracentral & 0.88 & 0.85 & 0.88 & 0.70 \\
parahippocampal & 0.82 & 0.74 & 0.80 & 0.61 \\
parsopercularis & 0.87 & 0.84 & 0.85 & 0.69 \\
parsorbitalis & 0.82 & 0.74 & 0.80 & 0.45 \\
parstriangularis & 0.86 & 0.81 & 0.85 & 0.71 \\
pericalcarine & 0.88 & 0.83 & 0.79 & 0.66 \\
postcentral & 0.92 & 0.87 & 0.90 & 0.77 \\
posteriorcingulate & 0.91 & 0.87 & 0.90 & 0.72 \\
precentral & 0.92 & 0.90 & 0.91 & 0.82 \\
precuneus & 0.93 & 0.87 & 0.91 & 0.80 \\
rostralanteriorcingulate & 0.78 & 0.75 & 0.79 & 0.23 \\
rostralmiddlefrontal & 0.91 & 0.88 & 0.90 & 0.78 \\
superiorfrontal & 0.92 & 0.91 & 0.92 & 0.78 \\
superiorparietal & 0.92 & 0.89 & 0.91 & 0.83 \\
superiortemporal & 0.90 & 0.91 & 0.91 & 0.68 \\
supramarginal & 0.91 & 0.87 & 0.90 & 0.70 \\
temporalpole & 0.43 & 0.64 & 0.62 & 0.37 \\
transversetemporal & 0.83 & 0.86 & 0.84 & 0.21 \\
\bottomrule
\end{tabular}
\end{table}

\begin{table}
\caption{Classification accuracy per cortical region of the Desikan-Killiany atlas \cite{desikan2006}. The results are computed using \qtyproduct[product-units = power]{1x1x5}{\milli\meter} \gls{flair} scans from the \gls{adni} GO-2 dataset. We use our model trained on synthetic scans (random contrast, random resolution). The reference, against which the accuracy is computed, is the default spherical registration pipeline from FreeSurfer.}
\label{tab:dice_region_flair_5mm}
\centering
\begin{tabular}{lSSSS}
\toprule
& {CAT} & {JOSA} & {ours (est)} & {ours (map)} \\
Region &  &  &  &  \\
\midrule
bankssts & 0.81 & 0.82 & 0.79 & 0.57 \\
caudalanteriorcingulate & 0.84 & 0.83 & 0.87 & 0.49 \\
caudalmiddlefrontal & 0.87 & 0.85 & 0.86 & 0.57 \\
cuneus & 0.87 & 0.75 & 0.82 & 0.70 \\
entorhinal & 0.67 & 0.64 & 0.72 & 0.47 \\
frontalpole & 0.59 & 0.44 & 0.54 & 0.27 \\
fusiform & 0.81 & 0.85 & 0.85 & 0.74 \\
inferiorparietal & 0.90 & 0.90 & 0.90 & 0.78 \\
inferiortemporal & 0.82 & 0.85 & 0.85 & 0.61 \\
insula & 0.91 & 0.92 & 0.92 & 0.79 \\
isthmuscingulate & 0.85 & 0.81 & 0.81 & 0.33 \\
lateraloccipital & 0.89 & 0.88 & 0.88 & 0.86 \\
lateralorbitofrontal & 0.93 & 0.90 & 0.93 & 0.80 \\
lingual & 0.88 & 0.88 & 0.88 & 0.80 \\
medialorbitofrontal & 0.90 & 0.87 & 0.90 & 0.73 \\
medialwall & 0.94 & 0.93 & 0.94 & 0.83 \\
middletemporal & 0.86 & 0.87 & 0.85 & 0.51 \\
paracentral & 0.90 & 0.86 & 0.88 & 0.77 \\
parahippocampal & 0.72 & 0.75 & 0.80 & 0.71 \\
parsopercularis & 0.87 & 0.84 & 0.85 & 0.72 \\
parsorbitalis & 0.82 & 0.76 & 0.81 & 0.59 \\
parstriangularis & 0.85 & 0.81 & 0.85 & 0.71 \\
pericalcarine & 0.86 & 0.81 & 0.81 & 0.66 \\
postcentral & 0.93 & 0.87 & 0.91 & 0.79 \\
posteriorcingulate & 0.87 & 0.89 & 0.88 & 0.66 \\
precentral & 0.93 & 0.90 & 0.92 & 0.77 \\
precuneus & 0.92 & 0.87 & 0.90 & 0.75 \\
rostralanteriorcingulate & 0.74 & 0.74 & 0.81 & 0.41 \\
rostralmiddlefrontal & 0.92 & 0.89 & 0.90 & 0.81 \\
superiorfrontal & 0.93 & 0.91 & 0.92 & 0.83 \\
superiorparietal & 0.92 & 0.90 & 0.91 & 0.80 \\
superiortemporal & 0.89 & 0.91 & 0.90 & 0.69 \\
supramarginal & 0.91 & 0.87 & 0.90 & 0.75 \\
temporalpole & 0.36 & 0.59 & 0.51 & 0.27 \\
transversetemporal & 0.79 & 0.85 & 0.83 & 0.46 \\
\bottomrule
\end{tabular}
\end{table}

\section{Discussion}

\subsection{The effect of resampling a surface onto the \emph{fsaverage} template}
\begin{figure}
    \centering
    \begin{subfigure}[t]{0.48\linewidth}
        \centering
        \includegraphics[width=\linewidth]{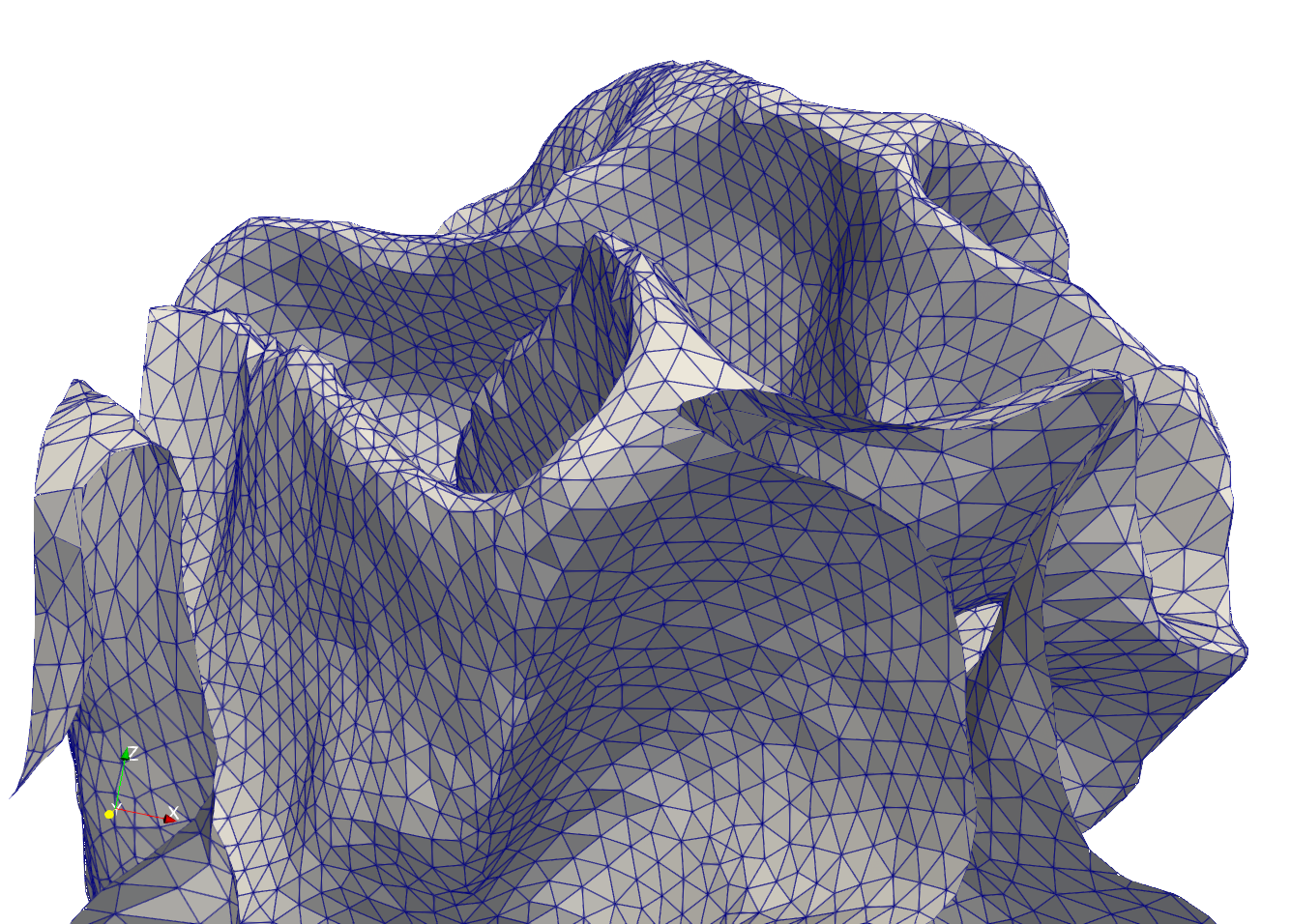}
        \caption{Original.}
        \label{fig:wm_surface_original}
    \end{subfigure}
    \begin{subfigure}[t]{0.48\linewidth}
        \centering
        \includegraphics[width=\linewidth]{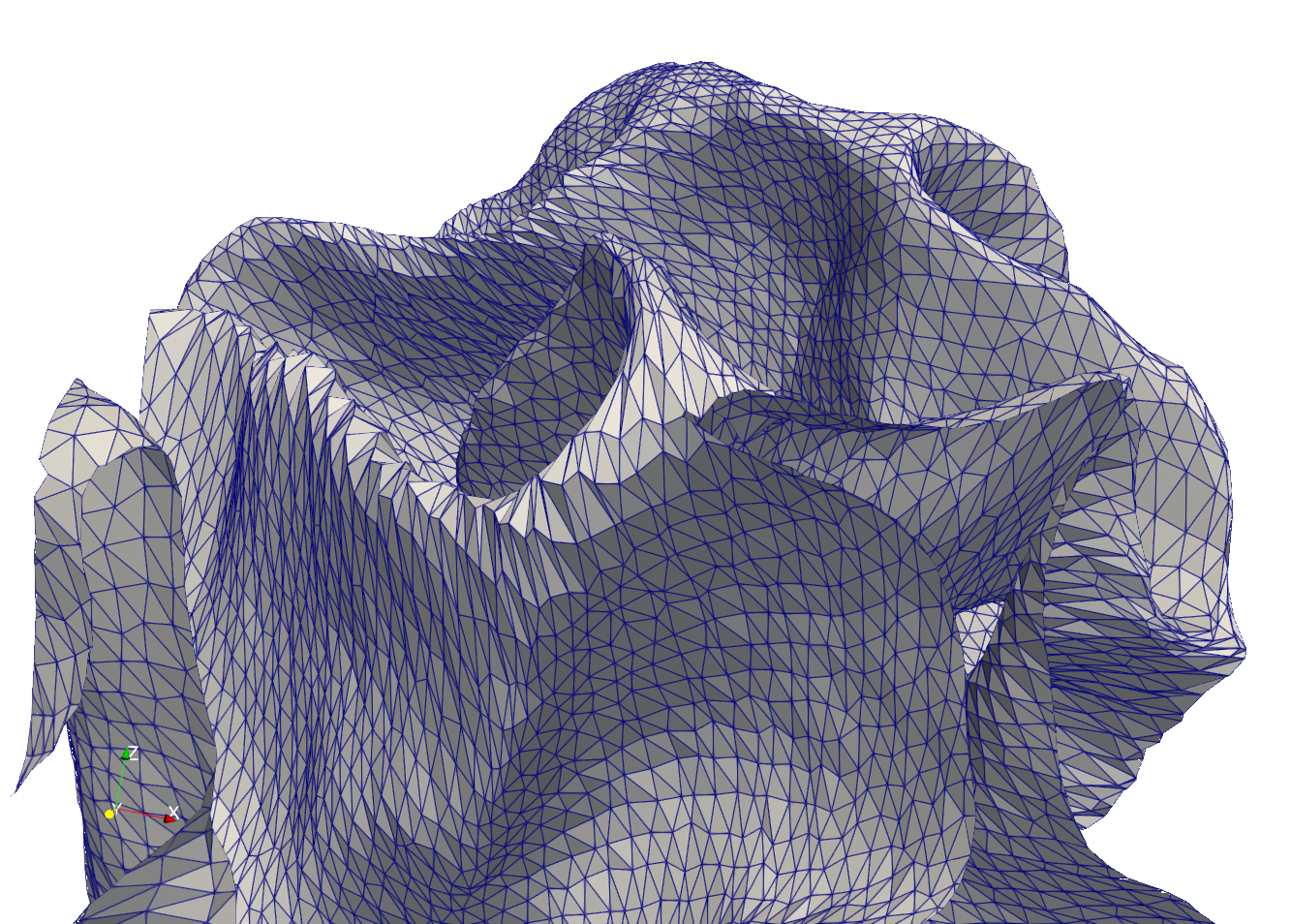}
        \caption{Resampled.}
        \label{fig:wm_surface_resampled}
    \end{subfigure}
    \caption{
    (\subref{fig:wm_surface_original}) Original white matter surface from FreeSurfer's \texttt{recon-all} pipeline.
    (\subref{fig:wm_surface_resampled}) White matter surface resampled to \emph{fsaverage}. Sheared triangles are clearly visible.
    }
    \label{fig:resampling_effect}
\end{figure}

In \cref{fig:resampling_effect}, we illustrate the negative effect of resampling a white matter surface (from FreeSurfer) to \emph{fsaverage} space. This clearly shows how sheared triangles are introduced. In the main text, we speculate that training exclusively with matched point loss on such surfaces (i.e., \cref{fig:wm_surface_resampled}) might have a negative effect on overall mesh quality.  

\bibliographystyle{plain}
\bibliography{references}

%% file: glossary.tex
\newacronym{abide}{ABIDE}{Autism Brain Imaging Data Exchange}
\newacronym{aibl}{AIBL}{Australian Imaging Biomarkers and Lifestyle Study of Aging}
\newacronym{adni}{ADNI}{Alzheimer's Disease Neuroimaging Initiative}
\newacronym{chinese-hcp}{Chinese-HCP}{Chinese HCP}
\newacronym{cobre}{COBRE}{Center for Biomedical Research  Excellence}
\newacronym{hcp}{HCP}{Human Connectome Project}
\newacronym{isbi2015}{ISBI2015}{2015 International Symposium in Biomedical Imaging}
\newacronym{mcic}{MCIC}{MIND Clinical Imaging Consortium}
\newacronym{oasis3}{OASIS3}{Open Access Series of Imaging Studies (phase 3)}
\newacronym{ixi}{IXI}{Information eXtraction from Images}
\newacronym{aomic}{AOMIC}{Amsterdam Open MRI Collection (ID1000)}

\newacronym{adhd}{ADHD}{attention-deficit/hyperactivity disorder}

\newacronym{gmm}{GMM}{Gaussian mixture model}
\newacronym{cnn}{CNN}{convolutional neural network}
\newacronym{gcn}{GCN}{graph convolutional neural network}
\newacronym{ode}{ODE}{ordinary differential equation}
\newacronym{prelu}{PReLU}{parametric rectified linear unit}
\newacronym{sdf}{SDF}{signed distance function}

\newacronym{ras}{RAS}{right-anterior-superior}

\newacronym{pv}{PV}{partial volume}

\newacronym{csf}{CSF}{cerebrospinal fluid}
\newacronym{gm}{GM}{gray matter}
\newacronym{wm}{WM}{white matter}

\newacronym{ct}{CT}{computerized tomography}
\newacronym{flash}{FLASH}{fast low-angle shot}
\newacronym{mr}{MR}{magnetic resonance}
\newacronym{mri}{MRI}{magnetic resonance imaging}
\newacronym{t1w}{T1w}{T1-weighted}
\newacronym{flair}{FLAIR}{fluid-attenuated inversion recovery}
\newacronym{mni}{MNI}{Montreal Neurological Institute}

\newacronym{charm}{CHARM}{Complete Head Anatomy Reconstruction Method}
\newacronym{cat12}{CAT12}{Computational Anatomy Toolbox 12}
\newacronym{rac}{\texttt{RAC}}{\texttt{recon-all-clinical}}
\newacronym{sif}{SIF}{self-intersecting faces}
\newacronym{dl}{DL}{deep learning}
\newacronym{assd}{ASSD}{average symmetric surface distance}
\newacronym{hd90}{HD90}{90 \% Hausdorff distance}